\documentclass[structabstract]{aa}  

\usepackage{graphicx}
\usepackage{txfonts}
\usepackage[]{natbib}

\usepackage{longtable,lscape}
\usepackage{subfig}

\bibpunct{(}{)}{;}{a}{}{,} 
\usepackage[dvips,usenames]{color}

\newcommand{\mytab}[6]{
\begin{table}
\caption{#4}
\label{#5}
\begin{center}
\begin{#6}
\begin{tabular}{#1}
\hline\hline 
#2
\hline
#3
\hline
\end{tabular}
\end{#6}
\end{center}
\end{table}
}

\newcommand{\myfig}[3]{
\begin{figure}
  \resizebox{\hsize}{!}{\includegraphics{#1}}
  \caption{#2}
  \label{#3}
\end{figure}
}

\usepackage{txfonts}

\newcommand{\mbol}{\ensuremath{M_{\rm{bol}}}}
\newcommand{\micron}{\ensuremath{\mu\rm{m}}}
\newcommand{\chir}{\ensuremath{\chi^2_{\rm{r}}}}

\begin{document}
\title{The VMC survey\thanks{Based on observations made with VISTA at ESO under program ID 179.B-2003.}}
\subtitle{III. Mass-loss rates and luminosities of LMC AGB stars}
\titlerunning{Mass-loss rates and luminosities of LMC AGB stars}

\author{M. Gullieuszik
  \inst{1}
  \and
  M. A. T. Groenewegen
  \inst{1}
  \and
  M.-R. L. Cioni
  \inst{2,3}\fnmsep\thanks{Research Fellow of the Alexander von Humboldt Foundation}
  \and
  R. de Grijs
  \inst{4,5}
  \and
  J. Th. van Loon
  \inst{6}
  \and
  L. Girardi
  \inst{7}
  \and
  V. D. Ivanov
  \inst{8}
  \and
  J. M. Oliveira
  \inst{6}
  \and
  J. P. Emerson
  \inst{9}
  \and
  R. Guandalini
  \inst{2}
}

\institute{Royal Observatory of Belgium, 
  Ringlaan 3, 1180 Brussel, Belgium
  \and 
  University of Hertforshire, 
  Physics Astronomy and Mathematics,
  Hatfield AL10 9AB, United Kingdom
  \and
  University Observatory Munich, 
  Scheinerstrasse 1, 81679 M\"unchen, Germany
  \and 
  Kavli Institute for Astronomy and Astrophysics, 
  Peking University, Yi He Yuan Lu 5, Hai Dian District, 
  Beijing 100871, China
  \and
  Department of Astronomy and Space Science, Kyung Hee University,
  Yongin-shi, Kyungki-do 449-701, Republic of Korea
  \and
    Astrophysics Group, Lennard-Jones Laboratories, 
  Keele University, Staffordshire ST5 5BG, UK 
  \and
  INAF, Osservatorio Astronomico di Padova, 
  Vicolo dell’Osservatorio 5, 35122 Padova, Italy 
  \and
  European Southern Observatory, 
  Av. Alonso de Córdoba 3107, Casilla 19, Santiago, Chile 
  \and
  Astronomy Unit, School of Physics \& Astronomy,
  Queen Mary, University of London, Mile End Road, London, E1 4NS, UK, 
}

\date{Received ???; accepted ???}


\abstract
{ Asymptotic Giant Branch (AGB) stars are major contributors to both
  the chemical enrichment of the interstellar medium and the
  integrated light of galaxies.  Despite its importance, the AGB is
  one of the least understood phases of stellar evolution. The main
  difficulties associated with detailed modelling of the AGB are
  related to the mass-loss process and the 3$^{\rm rd}$ dredge-up
  efficiency}
{ To provide direct measures of mass-loss rates and luminosities for a
  complete sample of AGB stars in the Large Magellanic Cloud,
  disentangling the C- and O-rich stellar populations.}
{ Dust radiative transfer models are presented for all 374 AGB stars
  candidates in one of the fields observed by the new VISTA survey of
  the Magellanic Clouds (VMC). Mass-loss rates, luminosities and a
  classification of C- and O-rich stars are derived by fitting the
  models to the spectral energy distribution (SED) obtained by
  combining VMC data with existing optical, near-, and mid-infrared
  photometry.  }
{The classification technique is reliable at a level of -- at worst --
  75\% and significantly better for the reddest dusty stars.  We
  classified none of the stars with a relevant mass-loss rate as
  O-rich, and we can exclude the presence of more than one dusty
  O-rich star at a $\sim94\%$ level.  The bolometric luminosity
  function we obtained is fully consistent with most of the literature
  data on the LMC and with the prediction of theoretical models, with
  a peak of the C-star distribution at $\mbol\simeq-4.8$ mag and no stars
  brighter than the classical AGB tip, at $\mbol=-7.1$ mag.  }
{This exploratory study shows that our method provides reliable
  mass-loss rates, luminosities and chemical classifications for all
  AGB stars.  These results offer already important constraints to AGB
  evolutionary models.  Most of our conclusions, especially for the
  rarer dust-enshrouded extreme AGB stars, are however strongly
  limited by the relatively small area covered by our
  study. Forthcoming VMC observations will easily remove this
  limitation.}

\maketitle
%

\section{Introduction}
The Asymptotic Giant Branch (AGB) is the last stage of active nuclear
burning for low and intermediate mass ($\sim $0.8-8 $M_\odot$) stars.
Although it is a short-lived phase
-- less than a few Myr \citep{gira+2010} -- 
AGB stars are major polluters of the
interstellar medium and major contributors to the integrated light of
galaxies \citep{renz+1986}. 
In spite of its relevance, the AGB is one of the least understood  phases
of stellar evolution, and a major source of disagreement between the
results of different population synthesis models.
This is mainly due to the difficulties associated with  modelling 
convective dredge-up and mass-loss processes.

Most of the available sets of evolutionary tracks and isochrones cover
AGB evolution in a very approximate way, ignoring crucial aspects of
the AGB evolution.  A more successful approach is based on synthetic
codes, in which the stellar evolution is described by means of
simplified relations derived from complete stellar models, while
convective dredge-up and mass loss are tuned by means of a few
adjustable parameters \citep{renz+1981,groe+1993,mari2002,cord+2007}.  This
approach provided the first database of stellar isochrones suitable
for reproduction of the basic observed properties of AGB stars, and in
particular the bright red tail of carbon stars \citep{mari+2008}.
Despite the great success of these models, there are still some
parameters (e.g., 3$^{\rm rd}$ dredge-up and mass-loss efficiency) that
need a detailed calibration, as shown by the disagreement between the
observed number and luminosity function (LF) of C- and O-rich stars in
nearby galaxies and model predictions
\citep{gull+2008,held+2010,gira+2010}.
A more detailed calibration of the models requires measures of
mass-loss rates and a reliable classification of C-rich and O-rich AGB
stellar populations in nearby stellar systems.

Direct measures of mass-loss rates in AGB stars are limited to samples
of a hundred stars in the Magellanic Clouds
\citep{vanl+2008,groe+2009}. Estimates using larger samples of stars are, to
date, based on empirical relations between mid-infrared (IR) colours
and mass-loss rates \citep{mats+2009}.  Recently, \cite{srin+2009}
used mid-IR photometry from the SAGE survey \citep{meix+2006} to
identify thousands of AGB star candidates in the LMC. Their spectral
energy distributions (SED) were fitted with dust-free photospheric
models for O- and C-rich stars. The mass-loss rates were then derived
from the excess in the fluxes observed in the 8 and 24 \micron\
bands. This was the first step of an ongoing project aimed at directly
measuring mass-loss rates for all their candidates using a grid of
dust shell models \citep[see][]{serg+2011}.

To calibrate the 3$^{\rm rd}$ dredge-up efficiency complete catalogues
of O- and C-rich AGB stars are needed. The most straightforward method
to obtain a reliable classification uses a spectroscopic
classification, but this would be extremely expensive in terms of
observing time. The most recent and impressive effort in this
direction is the SAGE-Spec \citep{wood+2011} spectroscopic survey,
which aims at a classification of mid-IR sources (mainly AGB stars) in
the LMC. The project was awarded 224.6 hr of Spitzer Space Telescope
observations, and provided a reliable classification of 197 point
sources.  This is clearly a minor fraction of the total LMC AGB
population, which consists of some tens of thousands of stars
\citep{cion+2006}.  A more efficient, although in principle less
reliable way to achieve a classification employs photometric data.
Different approaches have been adopted, mainly involving either
optical narrow-band or near-IR broad-band photometry
\citep[e.g.,][]{batt+2005,gull+2008,held+2010}.

The Magellanic Clouds (MC) are ideal targets to study the details of
AGB evolution, since they host a huge population of AGB stars covering
the whole age (i.e. initial mass) range and they are nearby galaxies,
easily accessible for observations of resolved stellar populations.
The immediate drawback of the proximity is the large sky area covered
by the MCs.  Only recently have some observational campaigns provided
coverage of a significant fraction of both the LMC and the SMC, in
particular:
 the optical Magellanic Clouds Photometric Survey
\citep[MCPS,][]{zari+2002,zari+2004};
the all-sky 2MASS \citep{skru+2006} and the MCs IRSF \citep{kato+2007} surveys
in the near-IR;
the SAGE-LMC \citep{meix+2006}, SAGE-SMC \citep{gord+2011}, S$^3$MC
\citep{bola+2007} Spiter surveys, and the AKARI/IRC LMC survey 
\citep{ita+2008} in the mid-IR.

The VISTA Magellanic Cloud survey \citep[VMC,][]{cion+2011} is one of
the six ESO public survey projects that are carried out with the new
ESO telescope VISTA \citep{emer+2010}.  It aims at imaging about 180 square degrees in the
Magellanic system (LMC, SMC, the Bridge and the Stream) in the $YJK_s$
wavebands, reaching a depth of $K_s\sim 20.5$ mag, nearly 5 mag
deeper than 2MASS and $\sim 3$ mag deeper than IRSF.
Observations started in November 2009 and will take about five years to reach
completion. The first images and catalogues were released to the
community in 2011 and new data will be released at regular intervals
according to the ESO policy for public surveys. The VMC data will
provide, e.g., a detailed history of star formation across the
Magellanic system and a measurement of its 3D geometry.  The VMC is
described in more detail in \cite{cion+2011}; a series of papers will present the
scientific results of the survey. To date, \cite{misz+2011} presented
a multi-wavelength study of LMC planetary nebulae and \cite{rube+2011}
obtained a spatially resolved Star Formation History of the LMC fields
already observed by the VMC.

We aim at extending the direct measure of luminosity and mass-loss
rate to the whole AGB stellar population in the Magellanic system by
modelling the SEDs obtained by collecting all available photometry
from optical to mid-IR wavelengths.  VMC observations are particularly
important, because their photometric quality and spatial resolution
are much higher than that of 2MASS. This yields a more accurate
photometry, in particular for the extremely red dust enshrouded AGB
stars, around the detection limit of 2MASS in the bluest bands --
$J\sim 16.5$ mag and $K_s\sim 15.0$ mag.  
In addition AGB stars are variables and the addition of VMC data
--taken at a different epoch from e.g. 2MASS-- will result in a more
robust estimate of the mean luminosity and mass-loss rate of the star.
The SED analysis also provides a classification of the stellar
chemical composition. To test its power to disentangle O- and C-rich
AGB stellar populations, we present a critical comparison between our
results and spectroscopic classifications from the literature.

In this paper we present our first results, based on
data of the LMC region covered by the first VMC observations.  The
data and the AGB selection criteria are presented in
Sect. \ref{sec:data}; Section \ref{sec:model} describes our model; the
results are presented in Sect. \ref{sec:results}; in
Sect. \ref{sec:summary} we summarise our work.

\section{The data}\label{sec:data}
We used the v1.0 VMC release, which is described in detail in
\cite{cion+2011}.
The data were processed by the VISTA Data Flow System \citep{emer+2004}
pipeline \citep{irwi+2004} and retrieved from the VISTA
Science Archive \citep{hamb+2004}.
The major issue for AGB stars in VMC photometry is saturation and
non-linearity of the detectors for bright objects. 
The VISTA  pipeline provides a correction
for saturation effects which can recover reliable photometry
for stars up to $\sim 2$ mag brighter than the saturation limit.
This is however not enough, since some AGB stars
are even brighter, so we had to reject VMC photometry
for stars brighter than 10.5 mag in the $Y$ and $J$ bands and 10.0 mag in the
$K_s$ band \citep[see][]{cion+2011}.

Additional photometry was obtained by collecting data from LMC 
photometric surveys:
\begin{itemize}
\item %
  optical: $UBVI$ data from the MCPS \citep{zari+2004};
\item %
  near-IR: $JHK_s$ data from the all-sky 2MASS \citep{skru+2006}, and
  the extended mission 6x long-exposure release; stars with very poor
  photometry -- quality flag ``E'' -- in any of the $JHK_s$ bands were discarded;
\item %
  mid-IR: IRAC 3.6, 4.5, 5.8, 8.0 and MIPS 24, 70, 160 \micron\ data from the
  SAGE catalogue \citep{meix+2006}; we used the
  \texttt{SAGELMCcatalogIRAC} and \texttt{SAGEcatalogMIPS24}
  catalogues containing both SAGE epochs 1 and 2 separately to avoid
  losing information on variability;
\item %
  mid-IR 9 and 18 \micron\ photometry from the AKARI/IRC mid-infrared
  all-sky survey \citep{ishi+2010}.

\end{itemize} 

The SED of each star was obtained using all
photometric data points. When a single magnitude
for each photometric band was needed, e.g., to plot colour-magnitude diagrams
(CMDs), the data were merged. For mid-IR photometry we used the original
SAGE epoch-merged data, while VMC and 2MASS
$J$ and $K_s$ were averaged after applying the colour
equations to transform $J$ and $K_s$ 2MASS photometry to the VISTA
photometric system \citep{cion+2011}.

For our project the use of optical and mid-IR photometry is
fundamental, so we will limit our analysis to the areas of the MCs
observed by all of the VMC, the MCPS and Spitzer. To date, there are only two VMC
fields in common with the optical and mid-IR surveys, namely fields
$6\_6$ and $8\_3$ \cite[see][]{cion+2011}. The first is centred on
the 30 Doradus star-forming region, which is affected by
severe crowding and differential reddening problems. Since in this
paper we want to define our methodology and test its reliability, it
is clear that data obtained from observations of field $6\_6$ are not
a suitable benchmark.  We consequently base the analysis presented in
this paper on data in VMC field $8\_3$, and in particular in the
1.42 deg$^2$ region delineated by 
$ 75 \fdg 50<\alpha\textrm{(J2000)}< 77 \fdg 50$
and
$-66 \fdg 95<\delta\textrm{(J2000)}<-65 \fdg 19$.

The targets for our analysis were selected using near- and mid-IR
photometry. 
We used 2MASS magnitude for selection rather than VMC ones, since some
bright AGB stars have no reliable VMC photometry because of
saturation.
The only problem with this choice could be those AGB stars that are heavily
obscured by thick circumstellar dusty discs with extremely red colours
that are too faint in the near-IR for detection by 2MASS. We verified
this hypothesis and checked that no such sources exist in the current sample.
The conditions we imposed are:
\begin{itemize}
\item %
  $J-K_s > -0.075 \times K_s+1.85$, the relation
  defined by \cite{cion+2006} to minimise contamination by
  foreground Milky Way and LMC red supergiant stellar populations.
\item %
  $K_s$ brighter than 12.0 mag, which corresponds to the tip of the RGB (TRGB)
  \citep{cion+2000}. 
  This condition was applied only to stars with $J-K_s <1.5$ mag, since
    redder stars cannot be RGB stars, and extremely red dust enshrouded AGB stars
  can be fainter than the $K_s$-band TRGB.
\item %
  IRAC 4.5 \micron\ magnitude brighter than 12.0 mag. This condition was
  set to avoid sources with faint mid-IR emission which cannot be AGB
  stars, but may be rather artifacts in the 2MASS catalogue.
\end{itemize} 
These criteria define a sample of 372 candidate AGB stars.  We
note that only 15 of them have 9 \micron\ AKARI photometry and only 4
are detected in the 18 \micron\ AKARI band. None are present in the
AKARI/FIS All-Sky Survey Bright Source Catalogue.  Finally, only one
of the selected sources was detected in the MIPS 160 \micron\ band and
three at 70 \micron.  All of these objects turned out to have SEDs
that are incompatible with AGB stars (see next section).

\begin{figure*}
  \centering
  \includegraphics[width=0.48\textwidth]{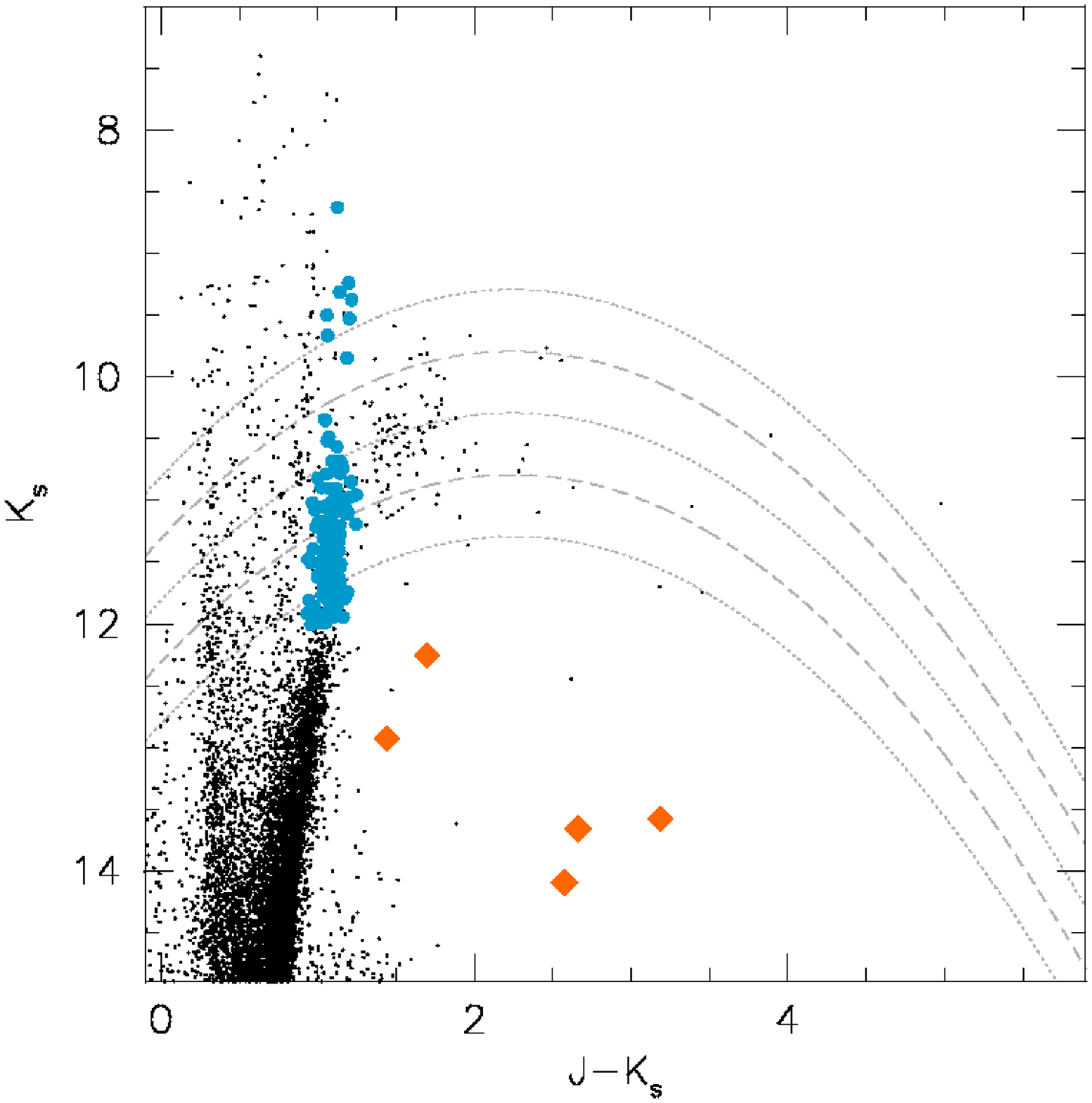}\hfill
  \includegraphics[width=0.48\textwidth]{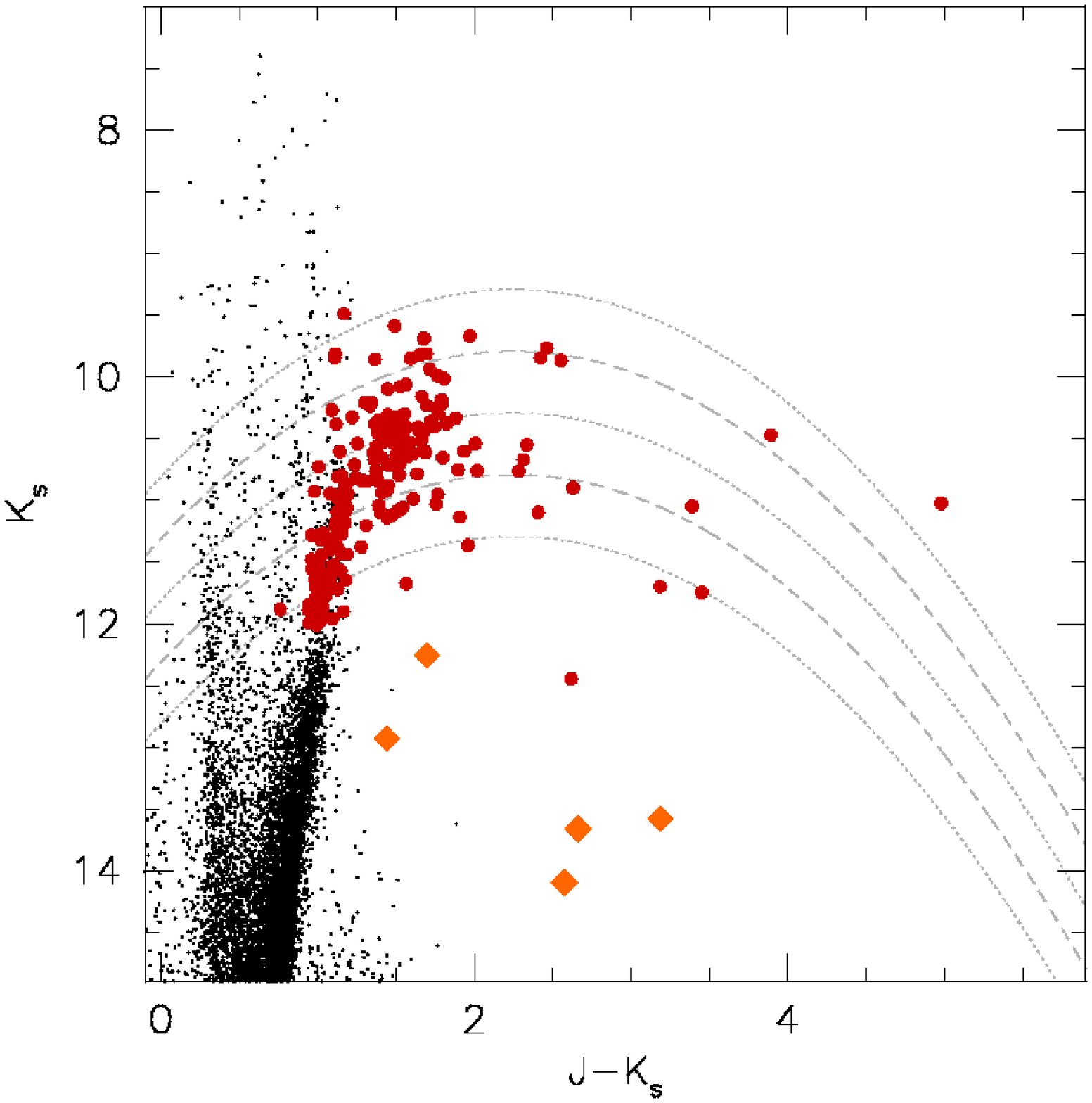}\\
  \includegraphics[width=0.48\textwidth]{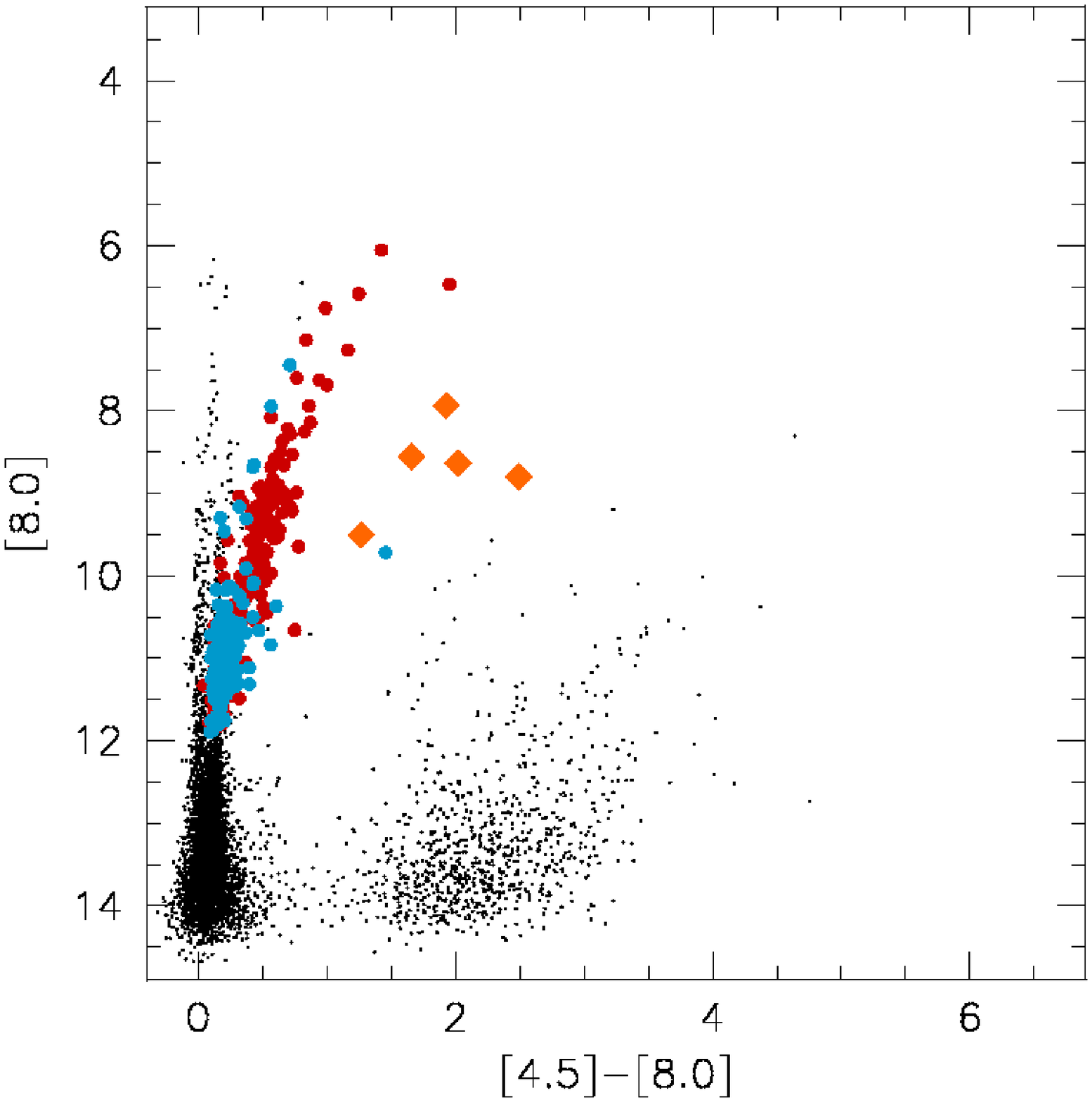}\hfill
  \includegraphics[width=0.48\textwidth]{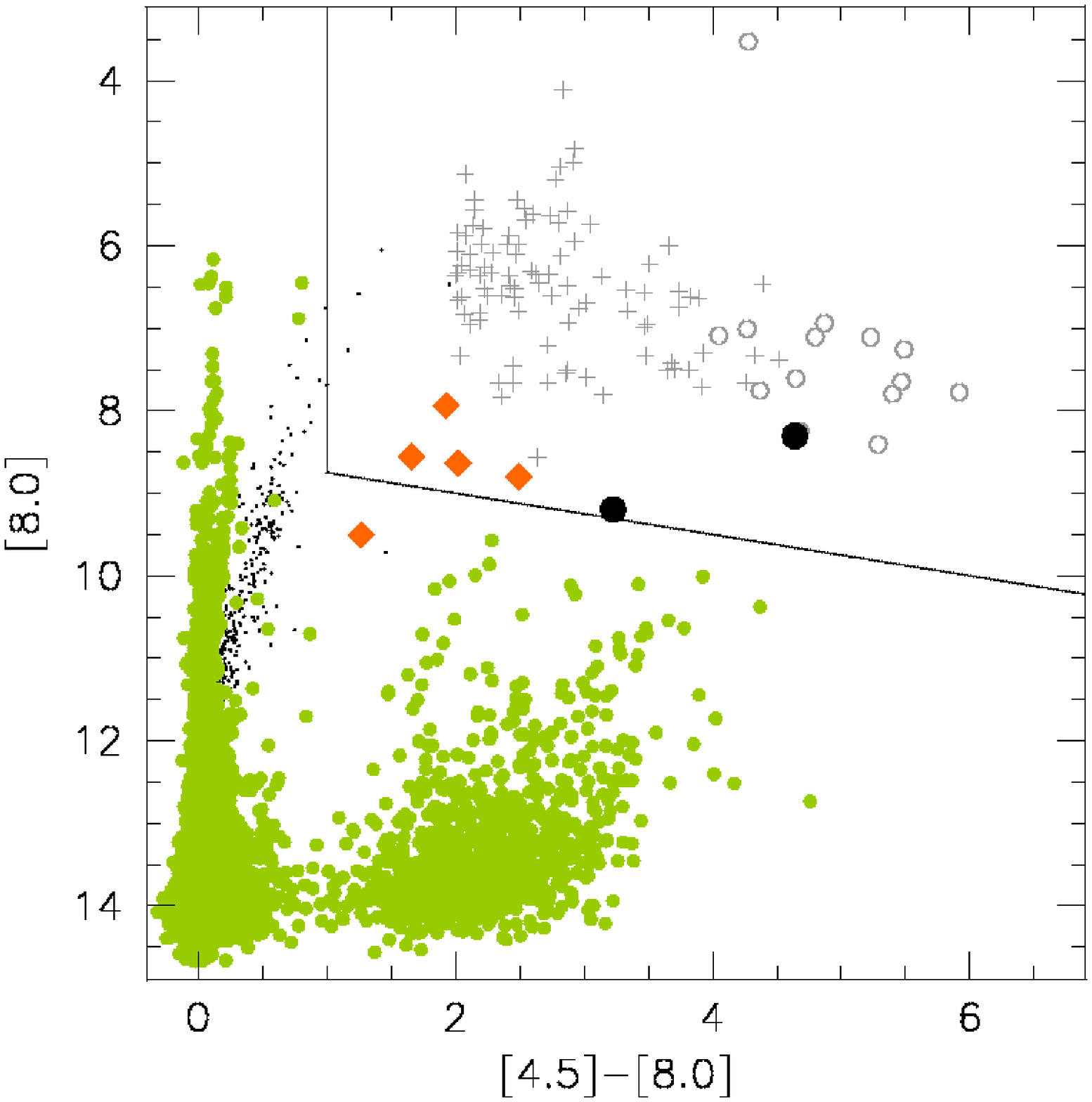}
  \caption{%
    Near- ({\it upper panels}) and mid-IR ({\it lower panels}) CMDs for stars in the VMC
    field 8\_3.
    Stars classified as O-rich {\it(upper left panel, blue dots)} and
    C-rich {\it(upper right panel, red dots)} are shown as solid symbols.
    The five non-AGB sources are displayed as diamonds.
    In the upper panels we plotted curves of constant bolometric
    magnitudes (\mbol\ between $-6.0$ and $-4.0$, with steps of 0.5 mag)
    derived using the $K_s$-band bolometric correction from
    \cite{kers+2010}.
    In the lower-left panel all AGB stars are shown as bigger points,
    while in the lower-right panel bigger (green) points show all stars not classified as AGB
    candidates using the near-IR criteria. The small grey crosses and open circles show all
    the sources in the LMC classified as AGB stars and EROs, respectively, by \cite{grue+2009}.
    The filled circles represent the two additional stars selected purely on the basis of the mid-IR photometry using the area delineated by the solid line.}
  \label{fig:cmdsel}
\end{figure*}

The selection criteria defined above could in principle exclude the most
extremely red AGB stars, like those discovered 
by \cite{grue+2008} using mid-IR photometry and spectroscopy.
Their SEDs peak in the mid-IR at wavelengths $\gtrsim 10\, \micron$
and they are extremely faint in bluer wave-bands.
Some of the stars in \cite{grue+2008} sample have IRAC 4.5 \micron\ magnitudes fainter than 12.0 mag,
and they could be likely fainter than the VMC detection limit.
For this reason, we added an additional selection criterion, based on
mid-IR photometry, to include extremely red stars in our sample. Using
all the stars classified as candidate AGB stars by \cite{grue+2009},
we defined the following selection rules: $[4.5]-[8.0] >1$ and $[8.0]
< 8.5+0.25 \times ([4.5]-[8.0])$, which correspond to the box
displayed in the lower-right panel of Fig. \ref{fig:cmdsel}. This
criterion adds two objects to our sample.  Anticipating the results
that will be presented in the next sections, we can safely exclude
that the bluest of the two objects is an AGB star. The classification
of second one is quite unclear, but in the Appendix of this paper we
provide some evidence that also this object could not be an AGB star.

\section{The model}\label{sec:model}

We fitted the SEDs with a combination of photospheric models, dust
models and by solving the radiative transfer equation.
The code used in this paper is based 
on that presented by \cite{groe+2009}, updated to
use the DUSTY code \citep{dusty} to solve the radiative
transfer equation (Groenewegen, in prep.).
Another important update is the use of \cite{arin+2009} models for
the stellar atmosphere of C-rich stars. For O-rich stars MARCS
models \citep{gust+2008} were used.

We adopted only two different dust-grain compositions, following
\cite{groe+2009}. A 94\% amorphous carbon + 6\% SiC mixture was assumed
for C-rich stars, and ``astronomical silicates'' with the absorption
coefficients of \cite{volk+1988} for O-rich stars.
  Since we are considering only broadband photometry,
we preferred to apply the results of
 \cite{groe+2009}, based on mid-IR spectroscopy,
rather than introducing the dust composition as a free parameter
weakly  constrained by the data used in our analysis.

The observed SED is fitted using a minimisation technique to find the
best-fitting values for the luminosity and optical depth $\vec{\tau}
_{\lambda}$ at 0.55 \micron\ of the dusty shell under the assumption
of a $r^{-2}$ density profile.  The minimisation is based on reduced
$\chi^2$ (\chir), which was calculated by taking into account the
photometric errors.  The mass-loss rate $\dot{M}$ was computed using
the relation of \cite{groe+1998}, applied to a geometrically thick
shell, since we assumed an outer to inner radius ratio of $10^4$
\citep{groe+2009}:

\begin{equation}
\vec{\tau} _{\lambda}=
5.405 \times 10^8
\frac{\dot{M} \, \Psi \, Q_\lambda/a}
{r_{\rm d} \, R_\star \, v_{\rm exp} \; \rho_{\rm d}} ,
\end{equation}
where 
$\dot{M}$ is in $M_{\odot}$ yr$^{-1}$, 
$v_{\rm exp}$ is the shell expansion velocity in km s$^{-1}$,
$R_\star$ is the stellar radius in solar radii, 
$r_{\rm d}$ is the inner dust radius in stellar radii,
$Q_{\lambda}$ is the absorption coefficient, 
$a$ is the dust grain radius in cm,
 $\Psi$ is the dust-to-gas mass ratio,
 and $\rho_{\rm d}$ is the grain density in g cm$^{-3}$.
 The values of $Q_\lambda/a$ and $R_\star$ were obtained from the dust
 and stellar models, respectively, while $r_{\rm d}$ is an output of
 the dust radiative code. We adopted $v_{\rm exp}=10$ km s$^{-1}$ and
 $\Psi=0.005$.  These are standard values for AGB stars in our Galaxy,
 which may not apply to the most metal-poor stars in the LMC \citep[see, e.g.,][]{vanl+2008}.

 The sample of stars was divided into  a ``Red'' and a
 ``Blue'' group, using as a discriminant the condition $J-K_s>1.5$ mag. 
The number of stars in the red and blue groups is 97 and 274, respectively.
The
 first group consists of the so-called red tail, the location in
 the CMD where most of the mass-losing C-stars are found. With this
 choice, in the ``Red'' group there are mostly low effective
 temperature dusty stars, whose SEDs are dominated by dust emission.
 In this case the choice of the atmospheric model is therefore not
 critical. On the other hand, bluer stars are expected to have higher
 $T_{\rm eff}$ and lower mass-loss rates and their SEDs are 
determined largely by the stellar photosphere rather than the dust envelope,
making the choice of photospheric model
 more critical. For the stars in this group we used a wider
 temperature range.
 Since the bluest stars are expected to have very low to negligible 
 mass-loss rates, for all temperatures we considered the possibility
 of having dust-free models, fixing 
 the mass-loss rate at zero. Only for the two lowest temperatures
 we left the mass-loss rate as a free parameter.
 These considerations lead to the definition of the adopted grid of
 atmospheric models, summarised in Table \ref{tab:modeltemp}.
 For the dusty models, we considered four different values for the
 condensation temperature of the dust grains ($T_c=$ 800, 900, 1000,
 and 1200 K).
In conclusion, for both groups we ran a set of 24 models; 
for the ``Red'' group we have 13 C-rich (two with four different $T_c$ plus
five dust-free) and 11 O-rich (two with four different $T_c$ plus three dust-free)
models;
for the ``Blue'' group we have 
12 (three with four different $T_c$) C-rich and 12 O-rich models.
For all stars we adopted a distance of 50 kpc \citep[e.g.,][]{scha2008}, an extinction $A_V=0.25$
mag \citep{schl+1998}, and the \cite{card+1989} extinction law.

\mytab{c|cc|cc|cc|cc} 
{
  $T_{\rm eff}$ {[}K{]}& 
  \multicolumn{4}{|c}{Blue}&
  \multicolumn{4}{|c}{Red}\\
  & 
  \multicolumn{2}{|c}{C-rich}&
  \multicolumn{2}{|c}{O-rich}& 
  \multicolumn{2}{|c}{C-rich}&
  \multicolumn{2}{|c}{O-rich}\\}{
  2600 &0&F& & & &F& & \\
  2800 & & & & & &F& &F \\
  3000 &0&F& & & &F& & \\
  3200 &0& &0&F& & & &F \\
  3600 &0& &0&F& & & &F \\
  4000 &0& &0& & & & & \\
} {The model grid.
  For each group (``Blue'' and ``Red'') and chemical composition,
  we list the temperatures adopted for models with
  the mass-loss rate set at zero (``0'') or as a free parameter 
  (``F'').} {tab:modeltemp} {center}

The model that yields the minimum \chir\  was taken as best-fitting
solution. Figure \ref{fig:seds} shows -- as an example -- the data and the 
best-fitting models for three stars.
The complete list of figures with the SEDs and the best-fitting
  models for all stars are available as online material, while the \chir\
values are listed in Table 6. The parameters of best-fit models
are presented in Table \ref{tab:datares}.
All stars with a \chir\ higher than 300 have been flagged as {\it B}
(``Bad fits'').  Most (e.g., the reddest star in Fig. \ref{fig:seds})
are extremely red stars, i.e. high-amplitude long-period
variables. The high \chir\ value can then be understood considering
that our data are obtained from observations carried out at different
epochs.  Seven sources however have SEDs incompatible with AGB stars
and they were therefore excluded from our analysis.  Four (IDs 73,
130, and 133 and 165 in Table \ref{tab:datares}) are classified as
candidate young stellar objects (YSO) \citep{wood+2011,grue+2009}, one
(ID 43) is a Seyfert galaxy at $z= 0.064$ \citep{vero+2010}, one (ID
14) is a post-AGB star \citep{wood+2011}.  The last object (ID 75), is
the source J050343.02-664456.7 in \cite{grue+2009} which is
generically classified as an Extremely Red Object (ERO). This is one
of the two objects not detected by VMC which were included in our
sample using the mid-IR selection criteria. The classification of this
source is doubtful. \cite{grue+2008} argued it is an AGB star, but in
the Appendix we show that its near-IR emission is unlikely to match
the SED expected for an AGB star. We suggest that it may be a
transition object, which left the AGB phase $\sim 100$ years ago, and
entered post-AGB phase.  Thus we excluded this star from our sample.
The final sample of objects with SED compatible with AGB stars is
therefore 367. The photometry for all of them is given in Table 6,
available as online material.

The photospheric model of the best-fitting solution gives also a useful
indication for classification of the stellar atmosphere as C- or
O-rich.  To estimate the confidence level of this classification, we
calculated the relative difference between the two \chir\
values of the two best-fitting solutions obtained considering O- and
C-rich models separately. A small difference indicates that the
solutions obtained with C- and O-rich models are equally acceptable;
if the difference is large, the solution with the lowest \chir\
gives a noticeably better description of the data, and in this case
the classification has a higher confidence level. After some tests, a
value of $\delta \chir=\Delta\chir / \chir=1$ was adopted as
discriminant.  Stars with $\delta \chir<1$ were flagged as {\it
  uncertain}.
Two tests as to the reliability of our classification -- based on
spectroscopic data from the literature-- are provided in
Sect. \ref{sect:reliabilityCO}.

\mytab{c@{\hspace{5pt}}c@{\hspace{5pt}}c@{\hspace{5pt}}c@{\hspace{5pt}}c@{\hspace{5pt}}c@{\hspace{5pt}}c@{\hspace{5pt}}c}{%
id&
RA (J2000)&
DEC (J2000)&
$\tau_{0.55}$&
$\log \dot M/M_{\sun}$&
$L/L_{\sun}$&
class&
flag\\
}{%
  1 &          05:02:00.86 &          $-$65:23:12.3 &        0.000 &      -99.999 &         3263 &      O &      U \\
  2 &          05:02:01.65 &          $-$66:16:00.2 &        0.000 &      -99.999 &         4847 &      O &      U \\
  3 &          05:02:02.88 &          $-$66:46:51.6 &        0.137 &       -7.462 &         5645 &      C &      S \\
  4 &          05:02:07.10 &          $-$66:56:03.1 &        0.019 &       -8.596 &         5402 &      O &      S \\
  5 &          05:02:09.01 &          $-$66:18:57.2 &        0.340 &       -7.069 &         5544 &      C &      S \\
}{%
  Parameters of best-fitting models for all sources in our database.
  Only the first entries are shown here as an example.  The full table
  is available at the CDS.
  The fourth column is the optical depth at $0.55\ \micron$.
  For stars without a dusty envelope, the logarithm of the
  mass-loss rate (5th column), was set to $-99.999$.  The 7th
  column indicates the C- and O-rich classification. Sources flagged
  ``B'' indicate stars with bad fits while ``U'' stands
  for uncertain classification and ``S'' for secure classification. 
 } {tab:datares}{center}

\myfig{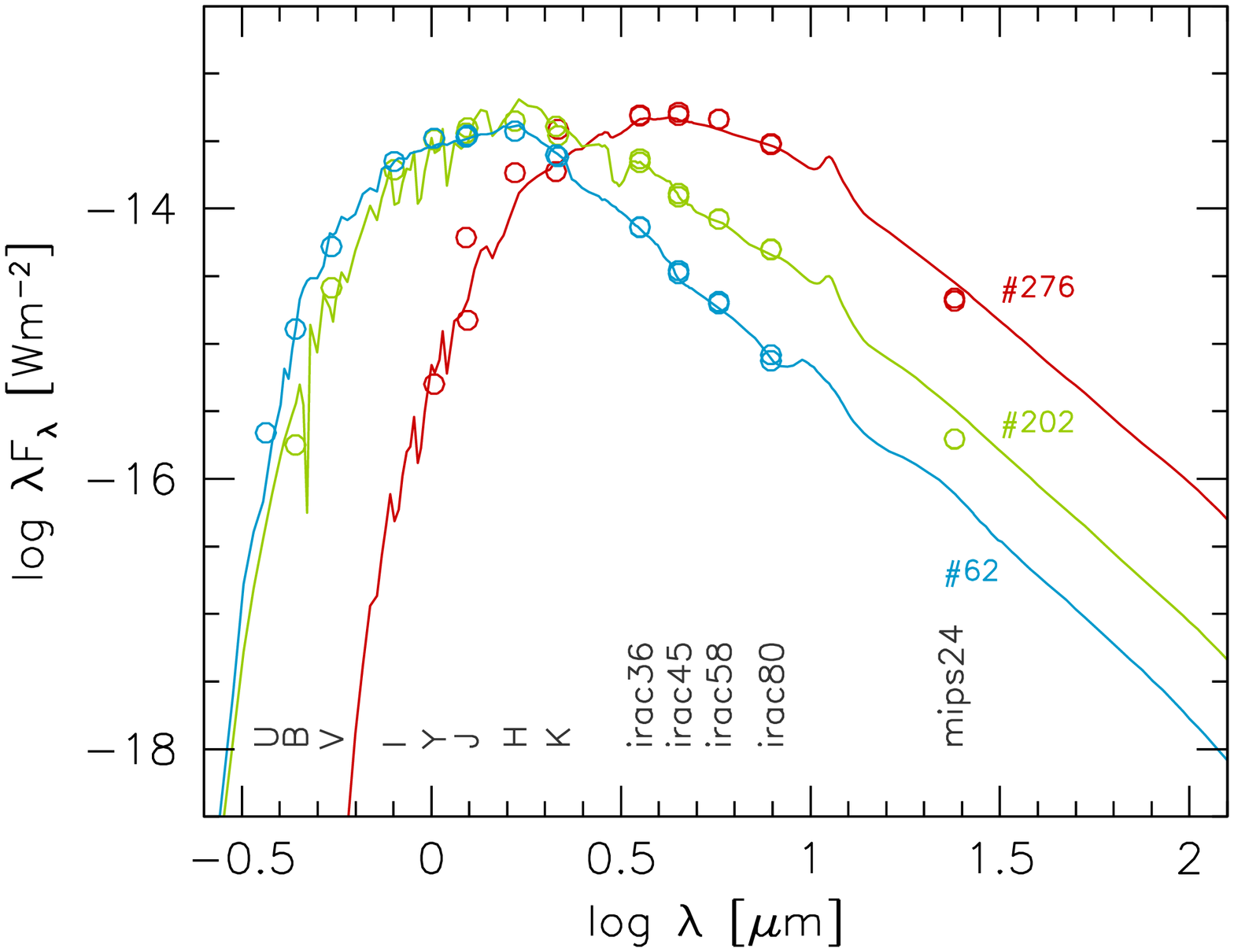}{
  Photometric data points ({\it open circles}) and the fitted SEDs
  ({\it lines}) for three AGB stars: a blue O-rich star ($J-K_s= 1.2$
  mag, id 62), and two C-rich stars ($J-K_s= 1.5$ mag, id 202; and $J-K_s
  \sim 4$ mag, id 276).
  }{fig:seds}

\begin{figure*}
  \centering
  \includegraphics[width=0.48\textwidth]{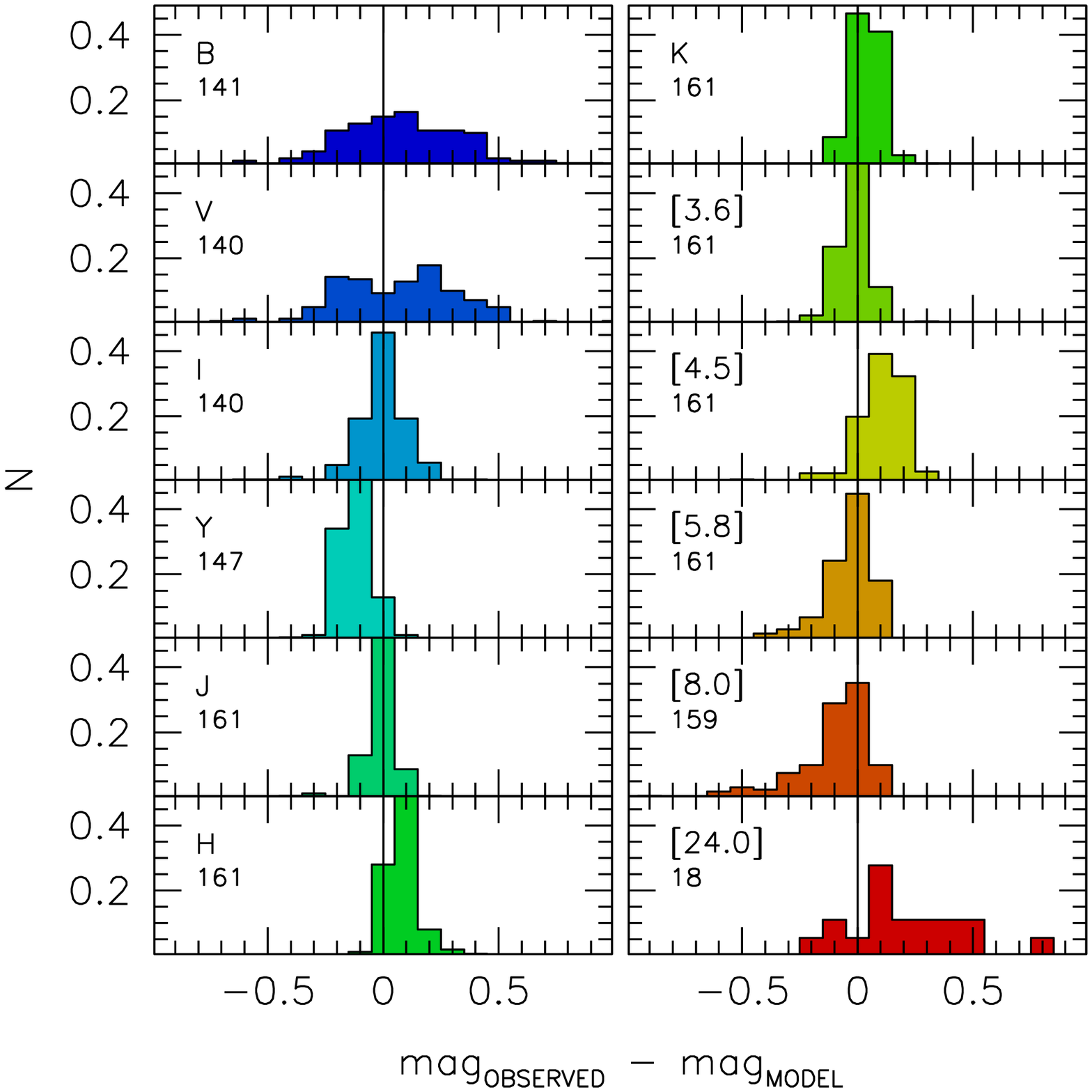}
  \includegraphics[width=0.48\textwidth]{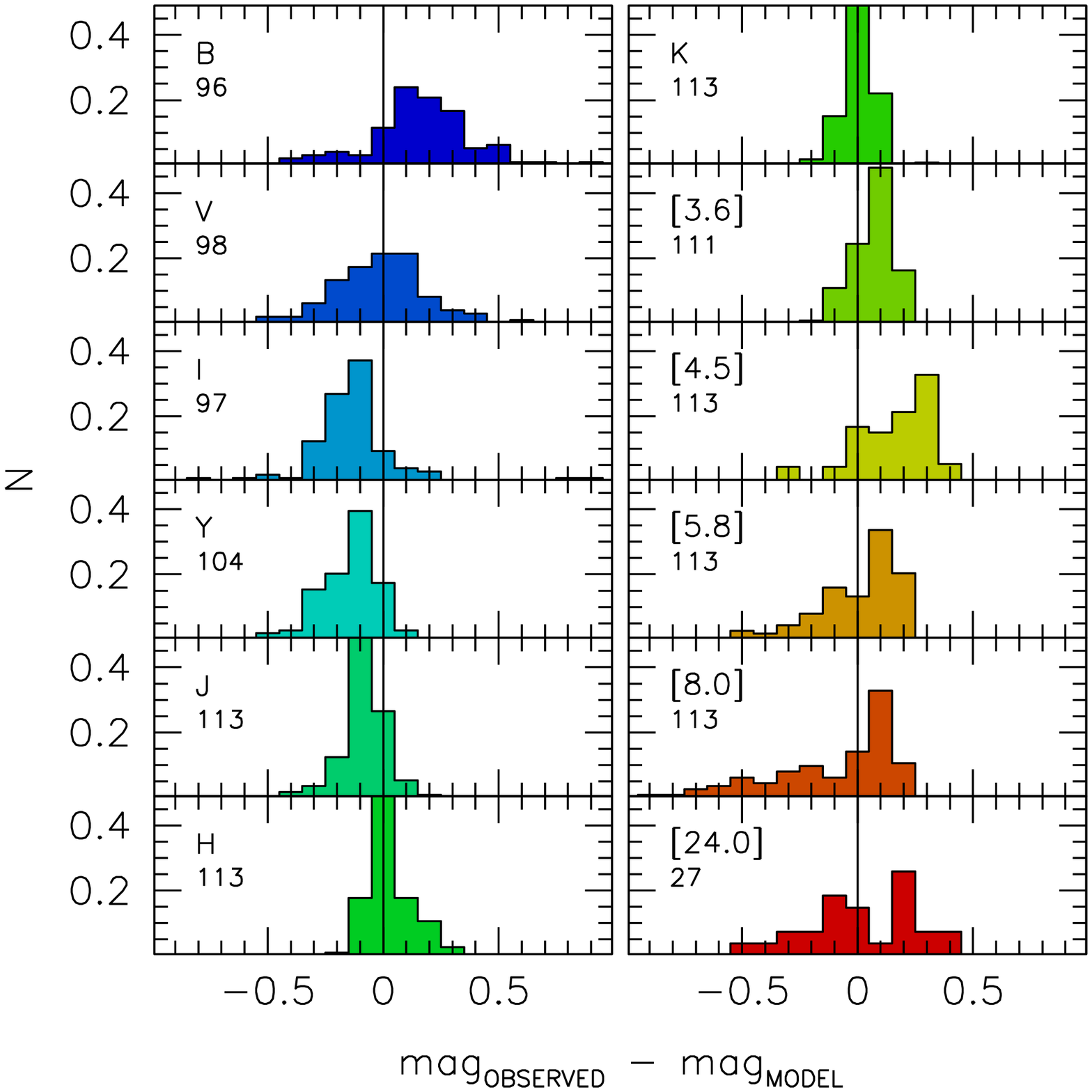}
  \caption{
    Histograms of magnitude residuals for the data in all wave-bands,
    for O-rich ({\it left panel}) and C-rich ({\it right panel}) stars
    with $J-K_s<1.5$ mag. The histograms are normalised to the total
    number of stars detected in each wave-band, labelled in each plot.
  }
  \label{fig:hfit_resblue}
\end{figure*}

    To check for any systematic effect in our modelling,
    Figs. \ref{fig:hfit_resblue} and
    \ref{fig:hfit_resred} show the differences between the observed
    magnitudes and those calculated from the best-fitting model in
    each photometric band.
  The variations of the distributions' spreads can be explained
  considering that AGB stars are variable sources, with amplitudes
  that are higher for redder stars and in the bluer optical bands.  At
  the bluest wavelengths ($B$ and $V$), the variation amplitude is
  maximum, as is the photometric error, since all stars --
  especially those in the red sample -- are faint (many red stars are
  not detected in the $B$ band).  On the other side of the spectrum,
  due to the limited sensitivity of the Spitzer Space Telescope in the
  reddest bands, most of the bluest stars are at the detection limit
  (only 45 out of the total of 274 blue stars are detected in the MIPS
  24 \micron\ band).
  The distribution of the residuals indicates a good quality of the fit
  for the stars in the blue sample, considering the photometric errors
  and the intrinsic variability of AGB stars.  The residual
  distributions are in general broader for stars in the red
  sample; this is not surprising, since these are the stars with
  the highest amplitude variability, where the effects of the
  circumstellar dust are more significant.  The scatter in the near-IR regime is
  similar in the red and in the blue samples, showing that the dust
  model -- which mostly affects the SEDs of the red stars -- is
  satisfactory.  Figure \ref{fig:hfit_resred} shows that there are
  systematic offsets in the reddest magnitudes. Models are always too
  faint in the IRAC 8 \micron\ and too bright in the MIPS 24 \micron\
  bands (see also Fig. \ref{fig:seds}). This effect is negligible
  for stars in the blue sample (Fig. \ref{fig:hfit_resblue}).
  The geometry of the dust shell, as well as the size and chemical
  composition of dust grains, influence the details of the dust
  emission, and different combinations of these parameters could in
  principle produce SEDs with different slopes at the red end.
  The systematic offsets in our models are however not completely
  surprising, since a similar effect can be seen also in
  \cite{groe+2009}. They used a similar SED fitting technique,
  but with the addition of Spitzer spectra, giving fundamental
  information on the dust emission features. They could therefore
  obtain much more information about the dust properties.
  In this paper we assumed the dust properties that provided the best
  description of the dust emission.  

\myfig{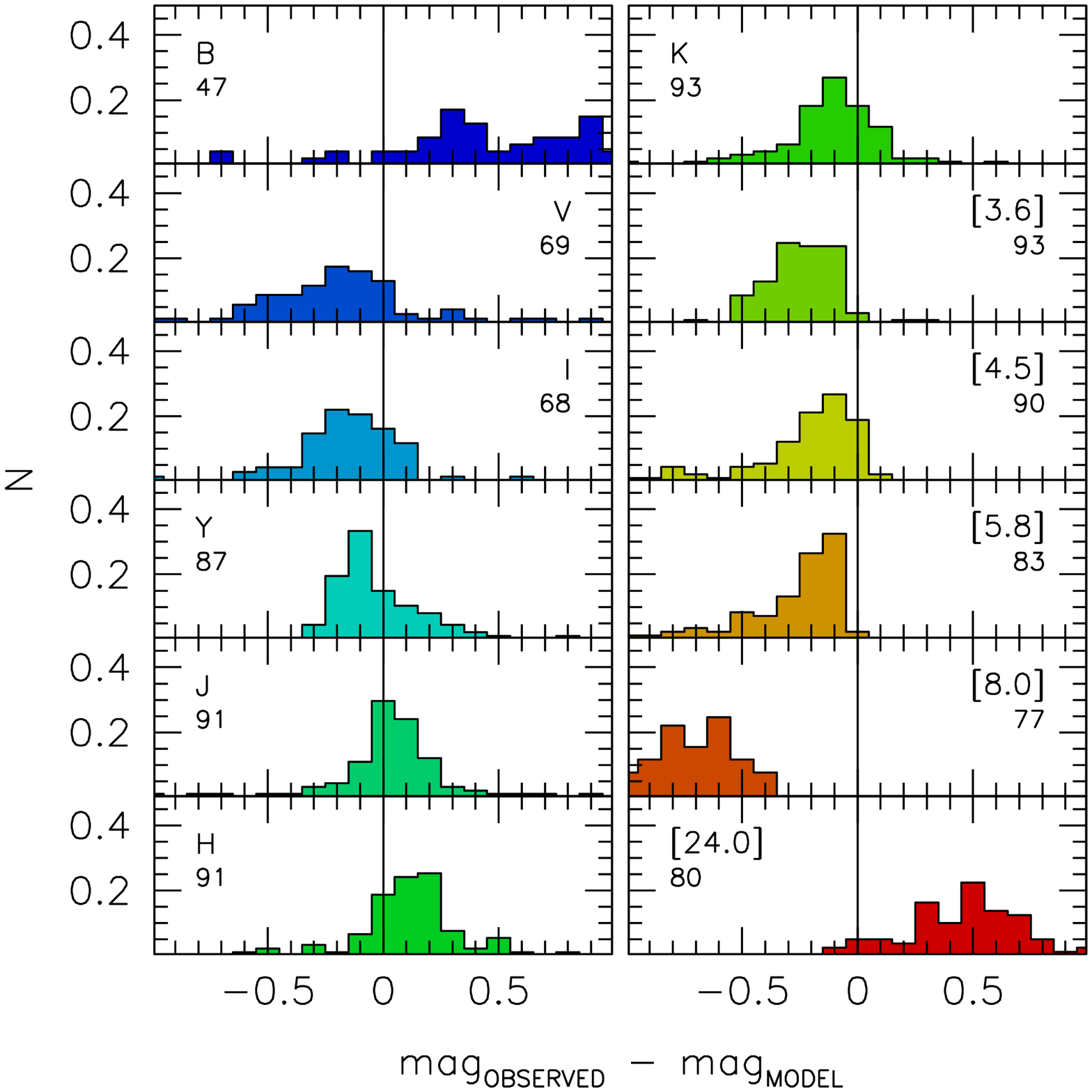}{
  Same as Fig. \ref{fig:hfit_resblue}, but for stars with $J-K_s>1.5$ mag.
  }{fig:hfit_resred}

\section{Results}\label{sec:results}

\subsection{Reliability of the classification}\label{sect:reliabilityCO}

To test the reliability of our C-/O-rich classification, we conducted two
different tests.  
First, we considered the catalogue of C-stars in the LMC
spectroscopically classified by \cite{kont+2001}.  We found 87 stars
in common with their sample and all are relatively blue
($J-K_s\lesssim2$ mag). Our procedure correctly classified as C-stars 87\%
of these. If we exclude the 20 stars with uncertain classification, this
fraction increases to 96\%.  All 54 stars with ($J-K_s>1.5$ mag) were
correctly classified, while the success rate for the 33 bluer ($J-K_s
\leq 1.5$ mag) stars is 67\%, or 84\% excluding the 14 stars with uncertain
classification.

Second, we considered the stars in the catalogue of \cite{groe+2009},
complemented with other LMC and SMC AGB stars with Spitzer IRS mid-IR
spectra (Groenewegen, in prep.\footnote{Compared 
  to \cite{groe+2009}, AGB and RSG stars from the following IRS
  programs were added:
30788 \citep[PI: Sahai , no dedicated publication, 3 spectra are discussed in][]{buch+2009};
40159 \citep[PI: Kemper, the SAGE-SPEC program described in][]{kemp+2010};
40650 \citep[PI: Looney, see ][]{grue+2008}; 
50167 (PI: Clayton,  no publication);
50240 \citep[PI: Sloan, the SMC-SPEC program, see ][]{sloa+2010}.
}).
The catalogue consists in 260 stars, and none of them
is located in VMC field 8\_3. The photometric data were therefore taken
from Groenewegen et al.
For homogeneity with our analysis, only data in optical, 2MASS and
Spitzer photometric bands were considered. We applied to the final
catalogue of 260 stars our fitting procedure and compared our
classification with that derived by Groenewegen et al. (2009, in
prep.).  The results are summarised in Table \ref{tab:confmartin}.
The reliability of our classification is confirmed to be extremely
high for C-stars.  In this case we obtained a $95\%$ confidence level
for classification of Blue C-rich stars, which is slightly higher than
the value obtained from the previous test based on the
\cite{kont+2001} results, but fully consistent with it considering the
low number of stars in the samples used for the two tests.  For O-rich
stars the confidence level is around $75\%$, similar as for red stars.

\mytab{ccccc}
{
  &
  \multicolumn{2}{c}{Blue}&
  \multicolumn{2}{c}{Red}\\
  &
  C&O&
  C&O\\
}
{
total
&  46  &  72  & 113   & 29\\
& (38) & (44) & (93)  &(22)\\[.2em]
correct
& 93\% & 72\% & 97\% & 76\% \\
& (95\%)& (82\%)&(100\%)& (73\%) \\ 
} {
  Results of the comparison with the classification by Groenewegen et
  al. (2009) and Groenewegen (in prep.).  Numbers in parentheses refer
  only to stars with robust classification (see text).
} {tab:confmartin} {center}

As an additional test, we considered the seven objects in the
SAGE-Spec sample \citep{wood+2011} located in the region used for our
study. Although this number is extremely small, the spectroscopic
SAGE-Spec classification is in perfect agreement with our results.
One of them is one of the YSOs and another one is the post-AGB star
 already discussed.
  Three sources
are red supergiants rightly excluded by the colour selection. Finally
two sources are classified as C-AGB stars both by \cite{wood+2011} and
by us.

\subsection{C- and O- rich stellar populations}

\mytab{ccccccc}
{
  & 
  \multicolumn{2}{c}{Blue}&
  \multicolumn{2}{c}{Red} &
  \multicolumn{2}{c}{All} \\
  & 
  C&O&
  C&O&
  C&O\\
}
{
Total                    & 113 & 161 &  93 &   0 & 206 & 161\\
certain classification   &  54 &  93 &  78 &   0 & 132 &  93\\
uncertain classification &  58 &  67 &  10 &   0 &  68 &  67\\
bad fit                  &   1 &   1 &   5 &   0 &   6 &   1\\

}
{Classification of AGB stars in the VMC 8\_3 field.}
{tab:class}
{center}

The results of our classification of the 367 AGB stars in VMC field
8\_3 are summarised in Table \ref{tab:class} and in the two upper
panels of Fig. \ref{fig:cmdsel}.
The location of C-and O-rich stars in the CMD agrees with predictions
of AGB stellar models \citep[e.g., see Fig. 7 in ][]{mari+2008}:
O-rich stars with no --or small-- dusty envelopes are located on a
nearly vertical blue plume, while C-rich stars populate the reddest
part of the CMD.  In \cite{mari+2008} this is ascribed to the cool
$T_{\rm eff}$ caused by changes in molecular opacities as the third
dredge-up events increase the photospheric C/O ratio.

The most remarkable fact is that all red stars are found to be
C-stars.  We cannot exclude, in principle, that some of them could be
misclassified O-rich. The results of our tests presented in
Sect. \ref{sect:reliabilityCO} showed that the probability of an
incorrect classification is around 25\%. The presence of more than one
O-rich dust-enshrouded star can therefore be excluded at a $\sim 94\%$
confidence level.

A very small number -- if any -- of red O-rich stars was expected a priori
since dust-enshrouded O-rich stars are rare:
low mass O-rich AGB stars do not reach high-mass rate (
$>10^{-5}M_\odot\,$yr$^{-1}$). On the other hand, the most massive
O-rich AGB stars are intrinsically rare objects, because of the shape
of the stellar initial mass function, which is dominated by low-mass
stars in a power-law fashion.  Dusty O-rich stars are therefore
expected to be found preferentially in regions with a high
star-formation rate at recent epochs ($<1$ Gyr).  The VMC 8\_3 field,
used for this preliminary study, is located in the outer regions of
the LMC, where the recent star-formation rate is not particularly high
\citep{rube+2011}.

All the brightest stars in the  blue vertical plume at $K_s<10$ mag 
are classified as O-rich, as expected \cite{mari+2008}. That region of the CMD is in fact
populated by the massive O-rich AGB stars in the hot bottom burning phase.
However note that we cannot exclude that some of them could 
in principle be core He-burning red supergiants.

\subsection{Bolometric magnitudes}

\myfig{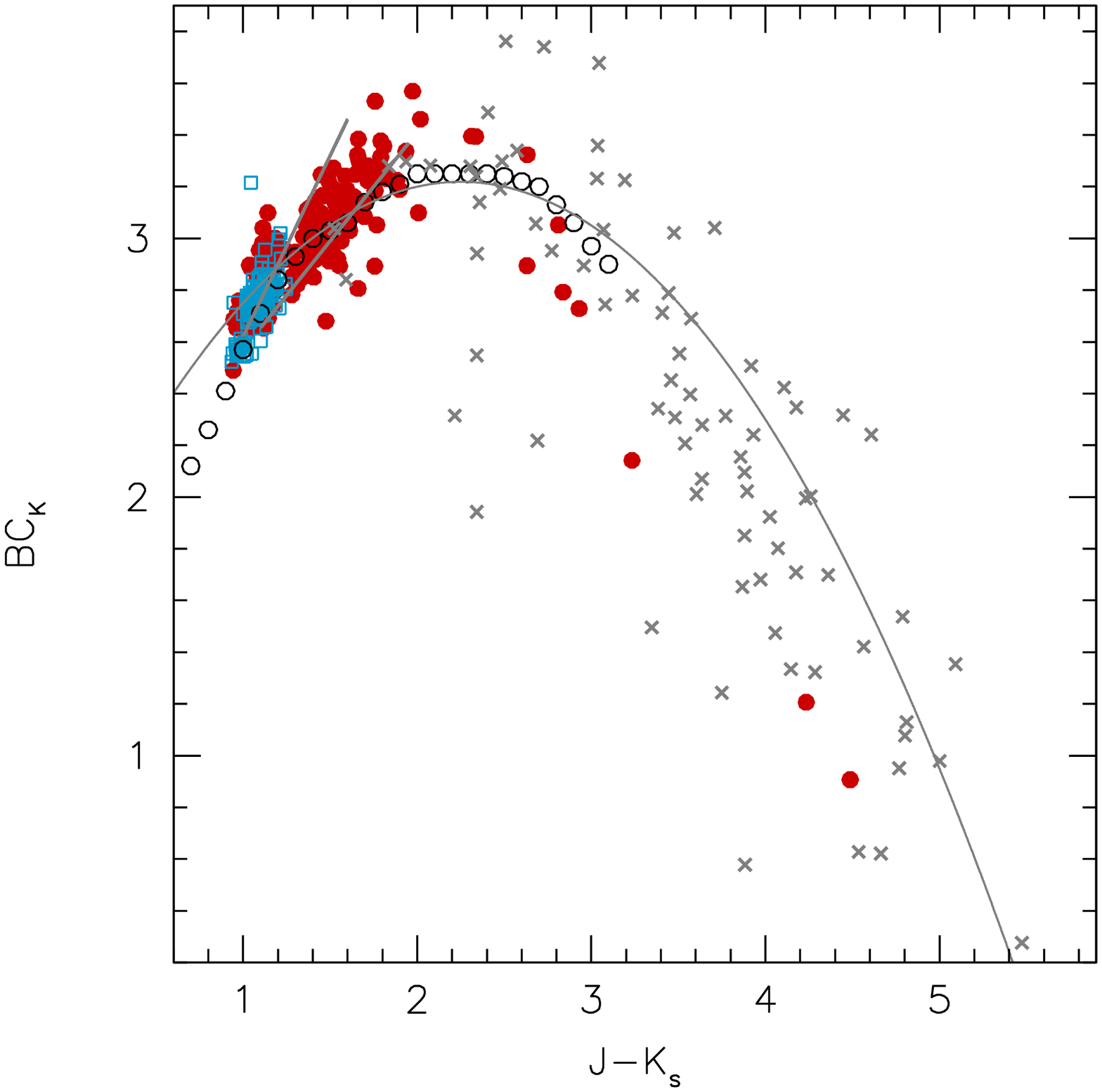}{Bolometric correction as a function of
  near-IR colour.  O-rich stars are shown as {\it blue open squares}
  and C-rich stars as {\it filled red circles}.
  {\it Crosses} are data for C-stars in the LMC and SMC from \cite{groe+2009}.
  The open black circles
  show the relation for Galactic C-stars from \cite{berg+2002}, the
  lines are taken from \cite{kers+2010}; the curved line is for
  C-stars, while the two straight lines define the region of O-rich stars.
}{fig:bcjk}

The bolometric correction $BC_K=\mbol-M_K$, obtained from the total
luminosity and the $K_s$-band magnitude (corrected for extinction and
distance modulus) is shown in Fig.  \ref{fig:bcjk} as a function of
$J-K_s$ colour. To extend our data points to redder colours, we
included also LMC and SMC data from \cite{groe+2009}.  Our data are
fully consistent with other literature data
\citep{berg+2002,kers+2010}. The only (possible) exception could be
represented by the extremely red C-stars at $J-K_s\gtrsim2.5$ mag. For
these stars the $BC$s obtained from both our data and those of
\cite{groe+2009} are located $0.3$ mag below the \cite{kers+2010}
relation. Note however that this relation is not well defined at
$J-K_s>2.5$ mag, since the sample of \cite{kers+2010}
contains only few red stars, which exhibit a significant scatter
and  lie -- on average -- below the best-fitting relation.
We can therefore conclude that the bolometric magnitudes obtained from
the SED fits are in agreement with literature data at a level of -- at
worst -- few tenths of magnitude.

\myfig{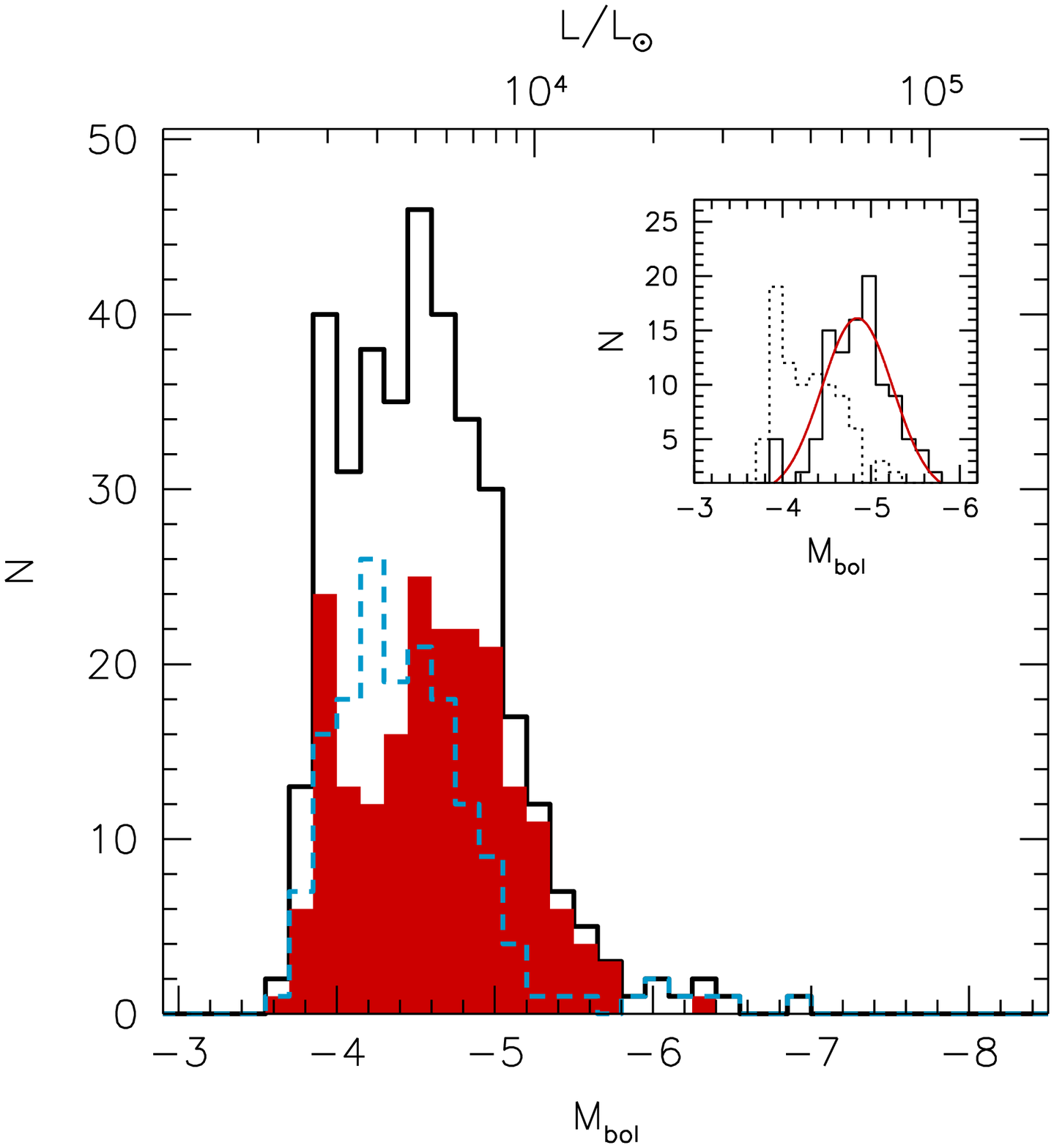}{Bolometric luminosity function of AGB stars.
  The {\it filled histogram} and the {\it dashed line} show the LF
  obtained from C-rich and O-rich stars, respectively.
  The inner panel shows the LF for dust-free ({\it dotted}) and
  dusty ({\it solid}) AGB C-stars. The solid curve is
  a Gaussian fit to the histogram of dusty stars.
 }{fig:lfbol}

 The bolometric LF function obtained for all AGB stars in our sample
 is shown in Fig. \ref{fig:lfbol}.  The cut at the faint end, around
 $\mbol=-3.7$ mag, is the consequence of the selection cut for stars
 fainter than the TRGB. The overall LF peaks at $\mbol=-4.5$ mag and
 drops at magnitudes brighter than $\mbol\sim-5.0$ mag.  There are
 very few stars brighter than $\mbol=-6.0$ and none above the
 classical limit for AGB stars, $\mbol=-7.1$ \citep{pacz1971},
 corresponding to a Chandrasekhar-limit core-mass before core He
 ignition.
 The most massive AGB stars, undergoing ``hot bottom burning'', may
 have a luminosity brighter than the classical AGB limit \citep[see,
 e.g.,][]{groe+2009,garc+2009}, but their number is expected to be
 extremely small.

 The LF of C-rich stars shows a bimodality, which can be understood
 by considering separately the LFs for stars whose best-fitting solution is
 dust-free and and stars with dusty envelopes. Dust-free stars show a
 decreasing LF, truncated by the selection cut.
   Dusty C-stars are typically brighter -- as already noted by \cite{vanl+1999} --
   and their LF is
 well described by a Gaussian function centred at $\mbol=-4.84$ mag
 and with $\sigma=0.40$ mag.  The peak of the LF is in perfect
 agreement with other observations of LMC C-stars and fully consistent
 with the expectations of theoretical models \cite[see,
 e.g.,][]{groe+1993}.  
 Some of the faintest AGB stars classified as C-rich could actually be
 misclassified O-rich stars, since our classification procedure is
 less precise for faint and blue stars, but in any case, our result
 seems to indicate the presence of a non negligible population of AGB
 stars fainter than the TRGB characterised by very low -- if any --
 mass-loss rates. This is in agreement with the prediction of recent
 evolutionary models, in contrast with the paucity of faint C stars in
 spectroscopic surveys \citep{kont+2001,gull+2008,mari+2008}.
 We will reserve a more detailed discussion of this issue for a future
 paper, presenting a comparison of our data with the predictions of
 theoretical evolutionary models.

 Recently, \cite{srin+2011} presented a C-star LF, revising their
 earlier work \citep{srin+2009}.  The general shape and the position
 of the peak in the revised LF is in good agreement with our result.

\subsection{Mass-loss rates}
\myfig{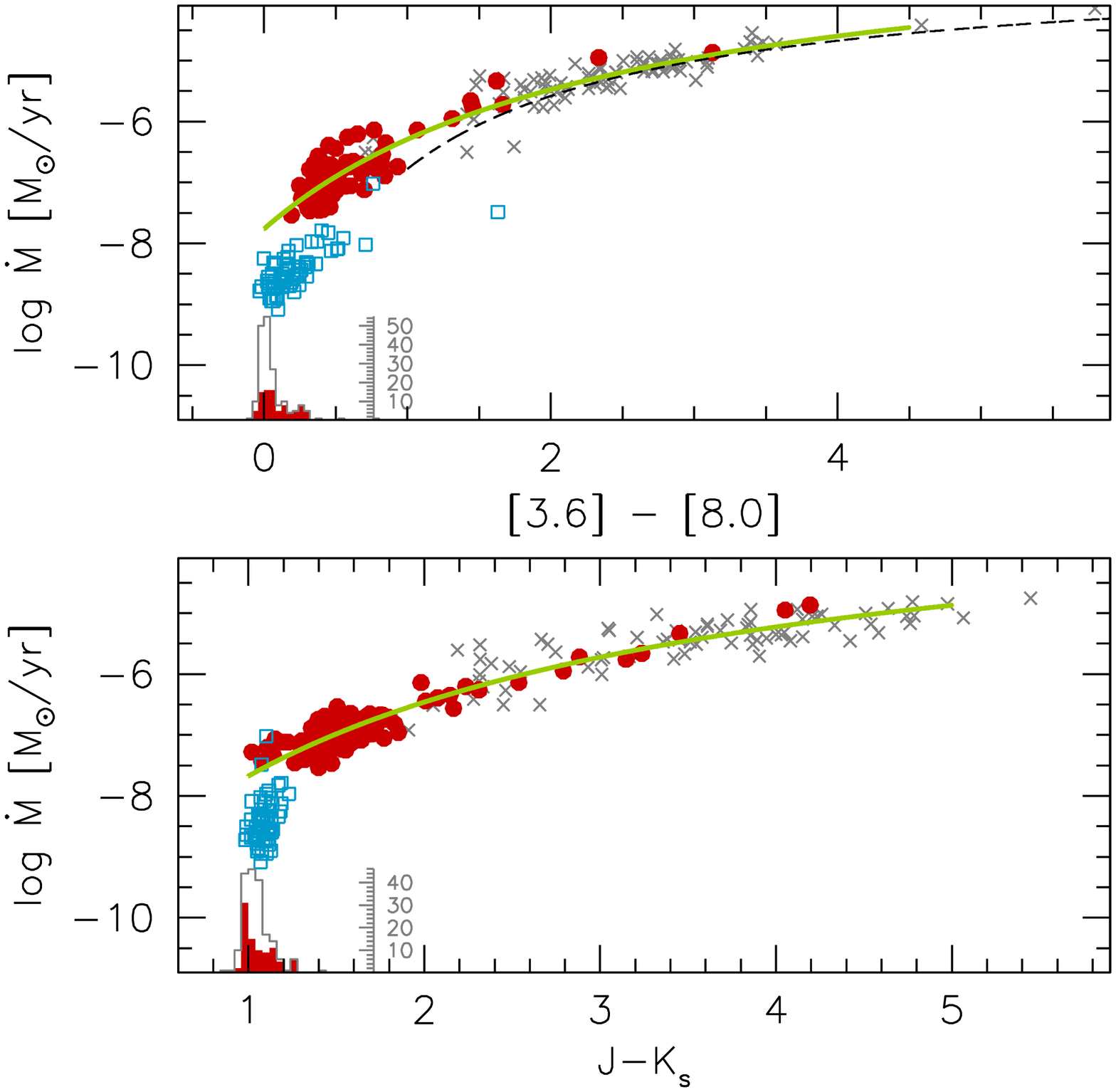}{
Mass-loss rates as a function of IR colours.
Symbols are the same as in Fig. \ref{fig:bcjk}.  The grey open
histogram shows the colour distribution of dust-free AGB stars; the
red histogram is the distribution of C-stars alone.  The solid line is the
best fit to all C-stars, including the \cite{groe+2009} data. The dashed line
in the upper panel is the relation of \cite{mats+2009}.
}{fig:mloss2P}

Figure \ref{fig:mloss2P} shows mass-loss rates as a function of IR
colours. The data points lie on a well defined sequence, fully
compatible with data published in \cite{groe+2009}, showing that our
fitting technique, based only on photometric observations, provides
reliable mass-loss rates.  Also in this case, the main limitation of
our results is due to the poor statistics for the reddest dusty
stars. We have no O-rich stars with noticeable mass-loss. The number
of dusty C-stars is still very small, but the sequence of the few data
points redder than $J-K_s\sim 2$ mag exhibits a remarkably low dispersion.
From a least-squares fit to all data, including those of
\cite{groe+2009}, we obtained the following relation for C-stars:
\begin{equation}\label{eq:mlossnir}
\log \dot M= \frac{-15.42}{(J-K_s)+ 2.10}-2.70
\end{equation}
for $1.0<J-K_s<5.0$ mag; and
\begin{equation}\label{eq:mlossmir}
\log \dot M=\frac{-12.78}{([3.6]-[8.0])+ 2.49}-2.63
\end{equation}
for $0.0<[3.6]-[8.0]<4.5$ mag.  For the bluest stars in our sample the
mass-loss rates are extremely low, at the detection limit of our
method.  At $J-K_s\sim 1.0$ mag, the best-fitting model was dust-free for most
stars; for the others, the mass-loss rates are less than
$\sim10^{-7}M_\odot\,$yr$^{-1}$. A conservative approach would suggest
that the measured values for the bluest stars must therefore be
considered upper limits.  The blue applicability limits of Eqs.
(\ref{eq:mlossnir}) and (\ref{eq:mlossmir}) were based on this
consideration and on the colour distributions for dust-free stars
in Fig. \ref{fig:mloss2P}. On the other hand, the definition of the
red limit is somewhat arbitrary and it was set by the colour of
the reddest stars in our sample.
The upper panel of Fig.  \ref{fig:mloss2P} compares our results with
the relation of \cite{mats+2009}, obtained using all measures
of \cite{groe+2009}. The difference for stars bluer than
[3.6]-[8.0]$=2$ mag is mainly due to the small number of stars in this
colour range in the \cite{groe+2009} sample. The relation of
\cite{mats+2009} is therefore weakly constrained for the bluest stars.
In addition, their relation is derived using the colours obtained
from the best-fitting SEDs, rather than the observed counterparts, as in our
case. The difference is therefore related to the systematic offsets
already discussed in Sect. \ref{sec:model}. We conclude that
our relation is more reliable when mass-loss rates are
estimated from observed photometry.

\section{Discussion and Conclusion}\label{sec:summary}

In this paper we used the first data from the ESO public survey VMC to
obtain optical to mid-IR SEDs for a sample of 367 AGB stars candidates.
Our selection criteria were defined to minimise the selection bias.
Sources with colours as blue as the RGB and fainter than $K_s=12.0$
mag --which roughly corresponds to the tip of the RGB in the LMC--
were excluded since this region of the CMD is dominated by the RGB and
it is nearly impossible to disentangle RGB and AGB stellar populations
from photometric data.

Luminosities and mass-loss rates were
obtained by fitting the SEDs with a dust radiative transfer model.  
Mass-loss rates were obtained assuming a constant expansion velocity and
dust-to-gas ratio with values similar to Galactic AGB stars.

\myfig{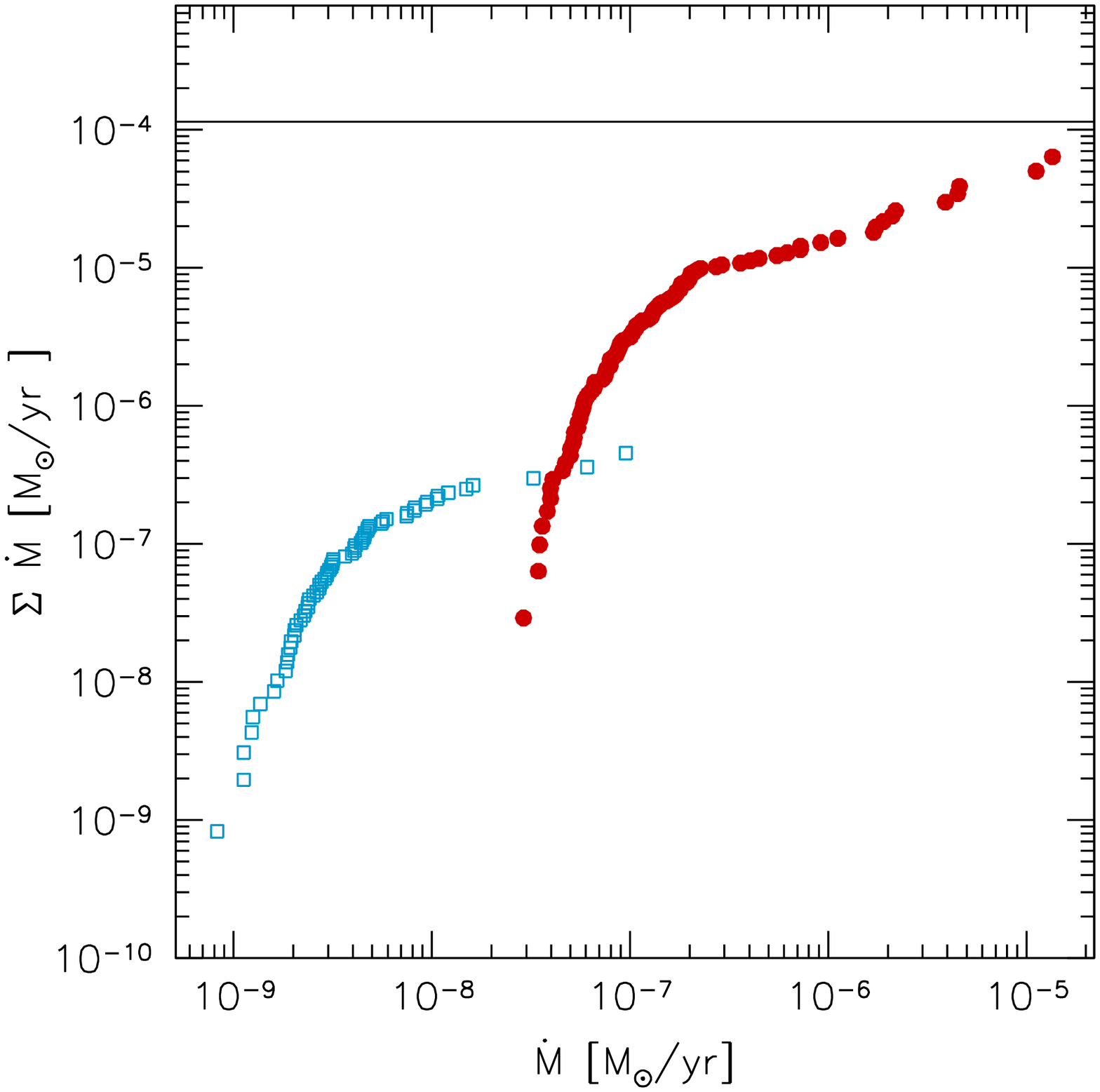}{Cumulative distribution of the mass-loss rates
for O- ({\it open squares}) and  C-rich ({\it filled dots}) AGB stars.
The horizontal line shows the average mass-loss rates of the external
regions of the LMC obtained by \cite{mats+2009}, rescaled to the area
encompassed by our analysis.}{fig:cumulml}

\begin{table}
  \caption{%
    Distribution of mass-losing O- and C-rich stars.
    The second column shows the star counts of \cite{mats+2009} for C-stars in
    the external region of the LMC
    rescaled to the area encompassed by our analysis.
    The last two columns are the star counts and total mass-loss rates 
    obtained from this study.}
\label{tab:mloss}
\begin{center}
\begin{tabular}{c@{ $<\dot{M}<$ }ccccc}
\hline\hline 
\multicolumn{5}{c}{C-rich}\\
\multicolumn{2}{c}{$\dot{M}$ range} & $N_{\rm{M09}}$ & $N$& $\dot{M}_{\rm{TOT}}$\\
\multicolumn{2}{c}{$M_\odot \ \rm{yr}^{-1}$} & & & $10^{-5} M_\odot \ \rm{yr}^{-1}$\\
\hline
\multicolumn{2}{c}{$< 1 \times 10^{-6}$} & 9.1 &102 & 1.5\\
$1 \times 10^{-6} $&$3 \times 10^{-6}$    & 8.1 &  6 & 1.1\\
$3 \times 10^{-6} $&$6 \times 10^{-6}$    & 4.4 &  3 & 1.3\\
$6 \times 10^{-6} $&$1 \times 10^{-5}$    & 1.8 &  0 & 0  \\
$1 \times 10^{-5} $&$3 \times 10^{-5}$    & 1.8 &  2 & 2.5\\
$3 \times 10^{-5} $&$6 \times 10^{-5}$    & 0.5 &  0 & 0 \\
\multicolumn{2}{c}{$> 6 \times 10^{-5}$} & 0.2 &  0 & 0  \\\hline
\multicolumn{2}{r}{Total:} & &  & 6.4\\\hline
\multicolumn{5}{c}{}\\
\multicolumn{5}{c}{O-rich}\\
\hline \hline
\multicolumn{2}{c}{$\dot{M}$ range}
 & $N_{\rm{M09}}$ & $N$& $\dot{M}_{\rm{TOT}}$\\\hline
\multicolumn{2}{c}{$< 1 \times 10^{-6}$} & - & 65 & 0.05\\\hline
\multicolumn{2}{r}{Total:} & &  & 0.05 \\
\hline
\end{tabular}
\end{center}
\end{table}

The cumulative distribution of mass-loss rates for O- and C-rich stars
is shown in Fig. \ref{fig:cumulml}. The total mass-loss rate in the
1.42 deg$^2$ area encompassed by our analysis is $6.4 \times 10^{-5}
\, M_\odot \, \rm{yr}^{-1}$. This value is lower than that expected
from the average mass-loss rate for the external regions of the LMC
found by \cite{mats+2009}, which is $1.1 \times 10^{-4} \, M_\odot \,
\rm{yr}^{-1}$ -- scaled to a 1.42 deg$^2$ area.  We note however that
the ERO object discussed in the Appendix was excluded from our
analysis.  \cite{mats+2009} considered it to be an AGB and estimated
for it $\dot{M}=6.4 \times 10^{-5} \, M_\odot \, \rm{yr}^{-1}$.
This object alone could explain the difference between the integrated
mass-loss rates.

Table \ref{tab:mloss} shows the star counts and the integrated
mass-loss rates for stars in different mass-loss ranges, together with
the star counts for the external region of the LMC from
\cite{mats+2009}.  Since \cite{mats+2009} considered only C-stars,
O-rich stars are not taken into account in Table \ref{tab:mloss}.  The
star counts are in good agreement, and the only significant
discrepancy is related to stars with low mass-loss rates. This is due
to the selection criteria adopted by \cite{mats+2009}, biased towards
red stars.  The 102 stars with $\dot{M}<10^{-6} \, M_\odot \,
\rm{yr}^{-1}$ contribute to $\sim 24\%$ of the total mass-loss rate,
and the two stars with higher mass-loss rates contribute up to 38 \%
of the total.  The fact that the integrated mass-loss rate is strongly
dominated by the most extreme stars, prevents us from drawing any
conclusion on the overall mass-loss rate in the LMC from the
relatively small number of stars in our sample. The highest values of
the mass-loss rates in the LMC can be as high as $\sim 10^{-4} \,
M_\odot \, \rm{yr}^{-1}$ \cite[e.g.,][]{grue+2008}.  To reach a more
robust estimate we therefore need to extend our analysis to a much
larger area.  The 65 mass-losing O-rich stars have an integrated
mass-loss rate of $4.5 \times 10^{-7} \, M_\odot\, \rm{yr}^{-1}$,
which is negligible compared to the C-stars.  Nevertheless, this is
not true in general, since extreme O-rich AGB stars can have mass-loss
rates as high as the integrated value of our sample
\citep{vanl+2005,groe+2009}.

The main limitations of the results reported in this paper are related
to the relatively small number of dust-enshrouded AGB stars in our
sample. The main aim of our work was to explore the confidence level
of the mass-loss rates and bolometric magnitude measures derived
from photometric SEDs constructed by combining VMC data with optical
MCPS, near-IR 2MASS and mid-IR Spitzer data, with a view to applying
our method to the complete sample of AGB stars that will be
observed by VMC in the
Magellanic system. 
We demonstrated the reliability of the measures and 
classification of the O- and C-rich AGB stellar populations.
A more complete picture about the mass-loss return and the process of
enrichment of the ISM will be obtained from forthcoming VMC observations.
VMC is scheduled to regularly carry out observations which, at the end of the 5 year
survey, will provide photometry for two orders of magnitude more AGB
stars.
Our results are nevertheless important, because our
database already contains a sufficiently large number of stars with
colours up to $J-K_s\simeq2$ mag, which represent the bulk of the AGB population.

\begin{acknowledgements}
  We would like to thank our referee, M. Matsuura, for her
  constructive comments which helped improve the quality of the paper.
  MG and MATG acknowledge financial support from the Belgian Federal
  Science Policy (project MO/33/026).
  RdG acknowledges partial research support through grant 11073001
  from the National Natural Science Foundation of China.
  %
  %
  The UK’s VISTA Data Flow System comprising the VISTA pipeline at the
  Cambridge Astronomical Survey Unit and the VISTA Science Archive at
  the Wide Field Astronomy Unit (Edinburgh) has been crucial in
  providing us with calibrated data products for this paper, and is
  supported by STFC.
\end{acknowledgements}
\bibliographystyle{aa} 
\bibliography{vmcagb} 
%
\begin{appendix}
\section{J050343.02-664456.7}

\begin{figure}
  \centering
  \resizebox{\columnwidth}{!}{\includegraphics{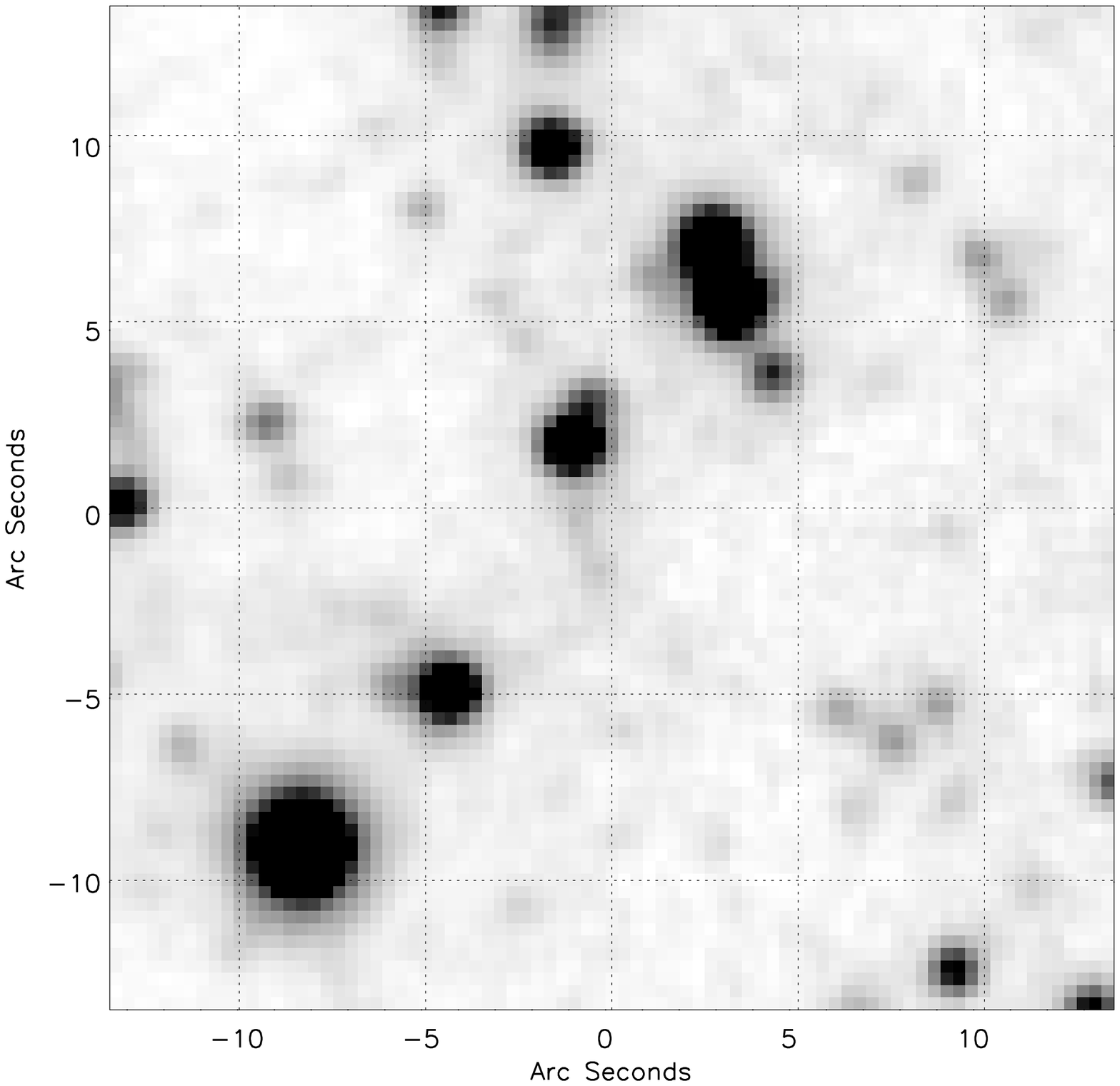}}\\
  \resizebox{\columnwidth}{!}{\includegraphics{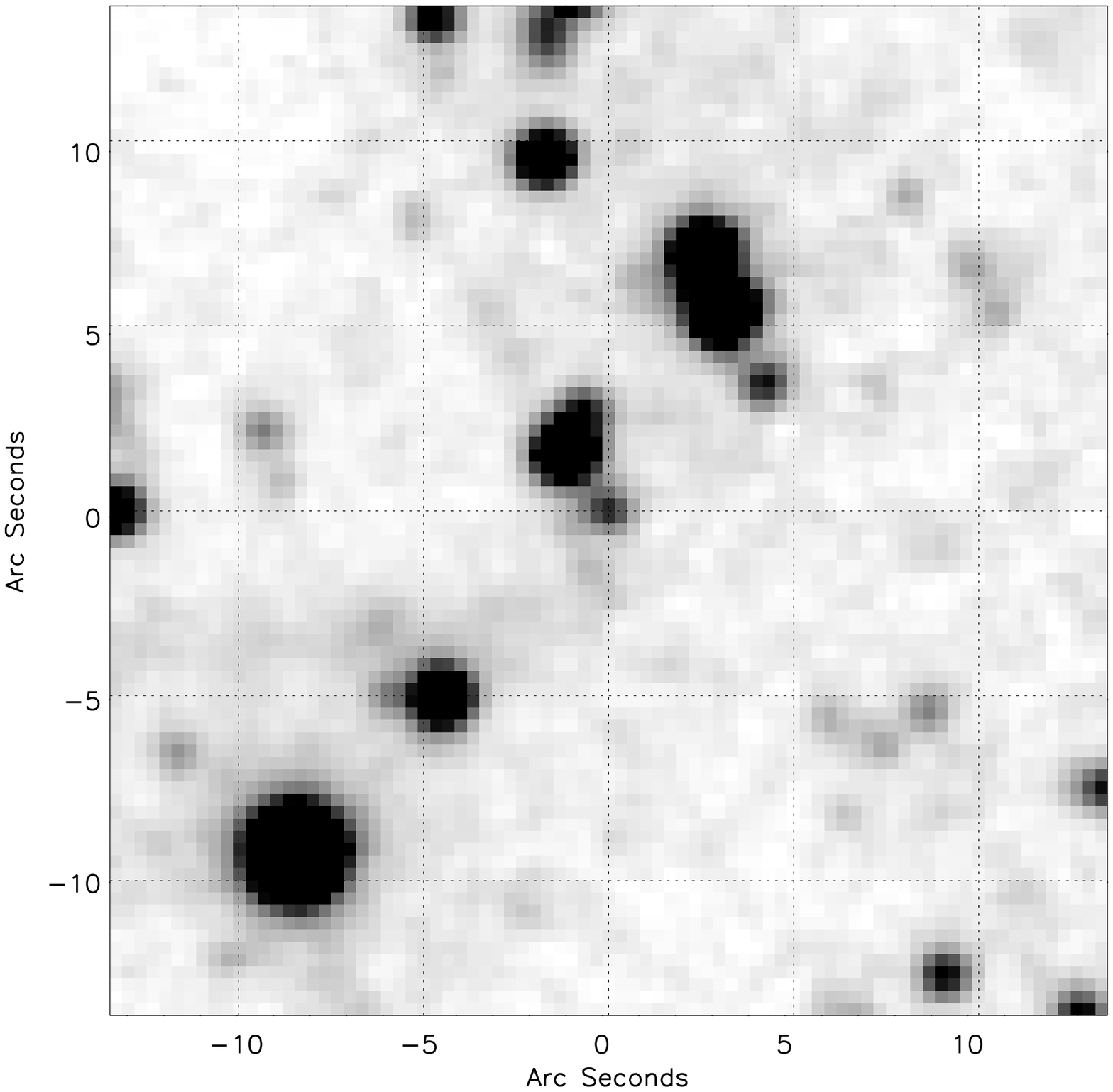}}
  \caption{
    $30\arcsec \times 30\arcsec$  $J$ ({\it upper panel}) and $K_s$ ({\it lower panel}) VMC images
    centred on the Extremely Red Object J050343.02-664456.
  }
  \label{fig:eroima}
\end{figure}

\begin{figure}
  \centering
  \resizebox{\columnwidth}{!}{\includegraphics{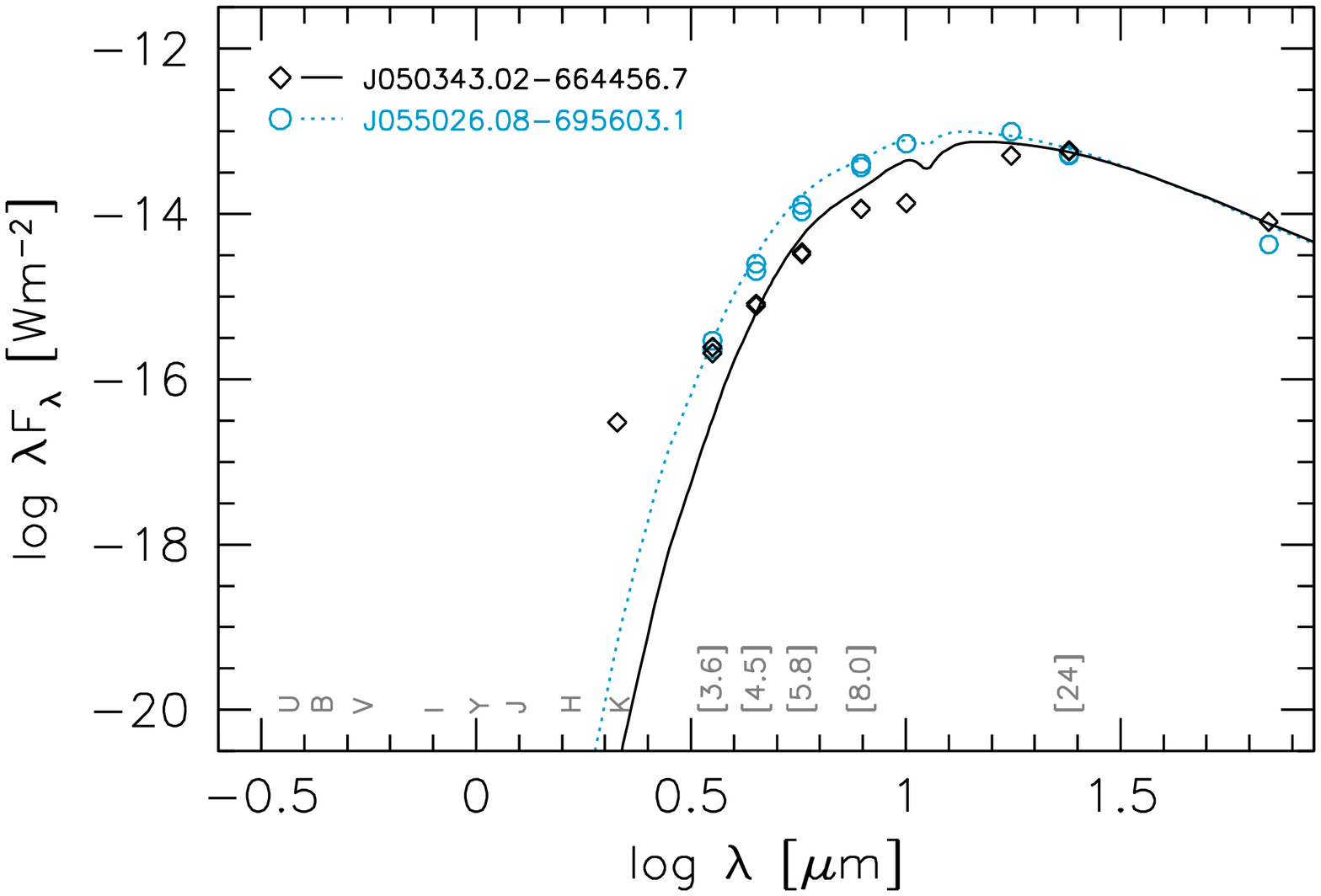}}\\
  \resizebox{\columnwidth}{!}{\includegraphics{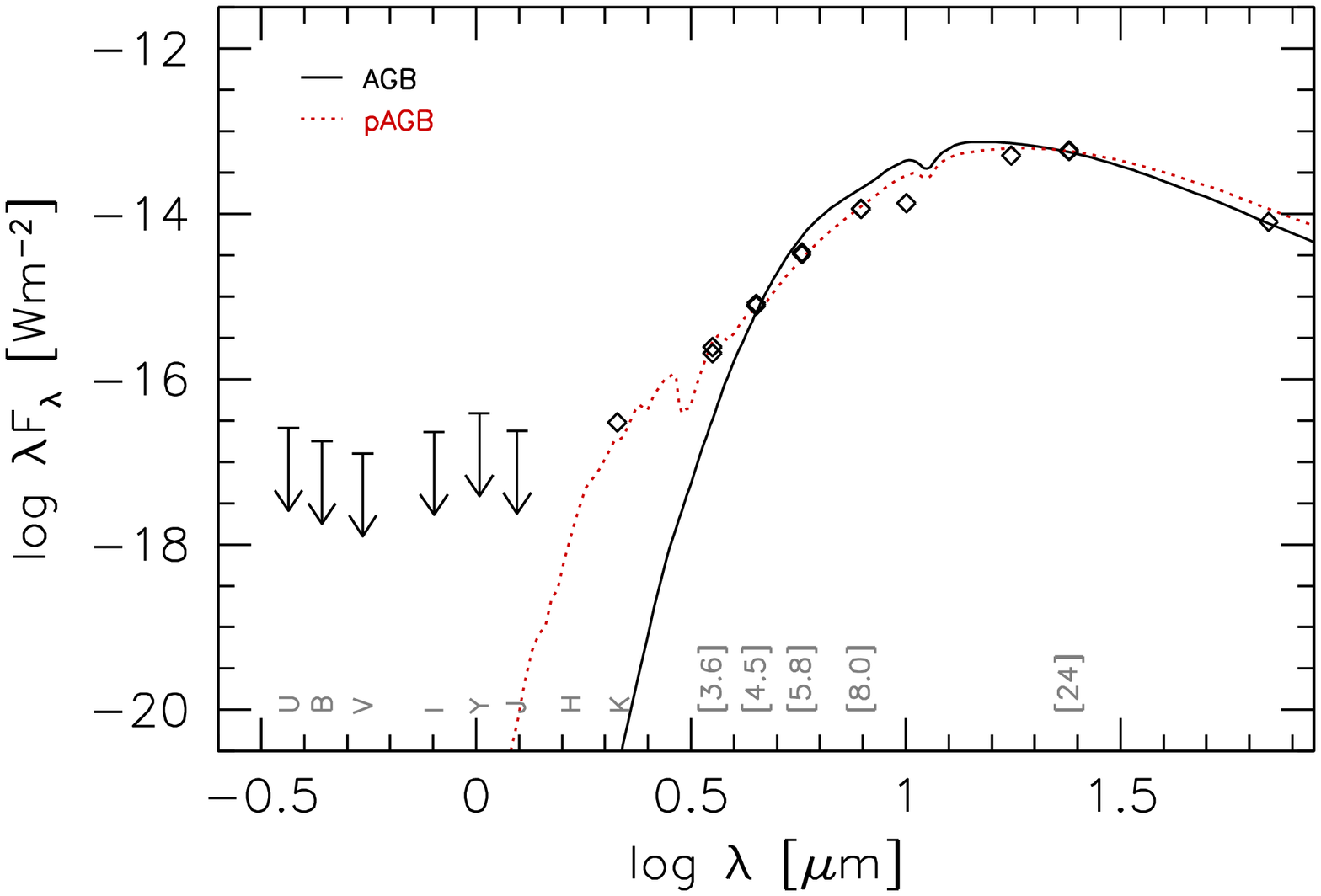}}
  \caption{
    Photometric data points and standard AGB SED model for 
    the Extremely Red Objects J050343.02-664456 ({\it diamonds and solid line}).
    In the upper panel they are compared with data and SED standard model for
    J055026.08-695603.1 ({\it (circles and dotted line}).
    The lower panel show the post-AGB model with a detached dusty shell
    {\it (dotted line}). The arrows show the photometric detection limits.
  }
  \label{fig:erosedcomp}
\end{figure}

This section is dedicated to a more detailed analysis of the extremely
red source J050343.02-664456.7. It is one of the reddest sources in
our sample, and one of two objects detected only in the mid-IR, with
no counterpart in the VMC catalogue -- the other one being a YSO.

J050343.02--664456.7 is one of the 13 sources identified by
\cite{grue+2009} and generically classified --on the basis of their
photometric SED-- as EROs.  Subsequent Spitzer/IRS spectroscopic
observations of seven of them revealed that they are extreme carbon
stars \citep{grue+2008}.  The conclusion is based on the detection of
SiC and C$_2$H$_2$ absorption and on the presence of a broad MgS feature
in two cases.

Assuming that these objects are AGB stars, the derived mass-loss rates
are  higher than  those of any known carbon-rich  AGB  star  in the  LMC.
\cite{grue+2008} estimated a mass-loss rate $\dot M=1.5 \times 10^{-4}
\,  M_\odot \,\rm{yr}^{-1}$  for J050343.02--664456.7.  From empirical
relations  based on  mid-IR colours,  \cite{mats+2009} found  a lower
value,  viz. $6.4  \times  10^{-5} \,  M_\odot \,\rm{yr}^{-1}$.   This
object  alone provides  a contribution  equal --or  bigger--  than the
integrated mass-loss  rate of all stars in our sample.
In   the   following   we  present   some   indications  that
J050343.02--664456.7 may not be a {\it normal} AGB star.

J050343.02--664456.7 is not present in the VMC catalogue because it is
detected only in the $K_s$-band.  It is in fact completely invisible
in the $J$ band as well as in the bluer $Y$ band, as shown in
Fig. \ref{fig:eroima}. The vmcSource catalogue is built with sources
detected in the three VMC bands \citep{cion+2011} and therefore
J050343.02-664456.7 is not included.  The $K_s$ magnitude was obtained
from the vmcDetection table, i.e. the catalogue corresponding to
individual observations. Its value is $K_s=18.75$ mag.

The photometric data points and the best-fitting model of
J050343.02-664456.7 obtained using the standard procedure described
in Sect. \ref{sec:model} are presented in
Fig. \ref{fig:erosedcomp}. The model substantially underestimates the
flux in the bluest part of the SED, in the $K_s$ and in the IRAC 3.6 \micron\ bands.
The predicted $K_s$ magnitude is $\sim 9$ mag
fainter than the observed one. 
We tested our model also on the other 
EROs  identified by \cite{grue+2009} --all located outside the $8\_3$ VMC tile and hence
not included in our sample-- 
 and found that in most cases
our AGB models could describe quite well the observed SED.
As an example, in
Fig. \ref{fig:erosedcomp} we show the photometric data and our
best-fitting SED model for J055026.08–695603.1, an EROs 
with a SED similar to the one of J050343.02-664456.7.
In this case the model SED seems to
generally better reproduce the observed data.
We are therefore led to consider the possibility 
that J050343.02-664456.7 could be something different from an AGB star.

The SED of J050343.02-664456.7 could be explained assuming that this
is a post-AGB star with a detached shell. In this case the flux in the
near-IR would be due to the contribution of the central star.  In our
case this contribution is extremely low, and produces only a flatter
SED in the near-IR, rather than a secondary peak in the near-IR
\citep[see, e.g.,][]{laga+2011}.
This implies that the central star stopped losing mass --i.e. it ended the AGB phase--
very recently \cite[see, e.g.,][]{vand+1989}.

To explore the post-AGB hypothesis,
we used our models setting the condensation temperature of the 
dust grains as a free parameter. The resulting best-fitting model  is
shown in the lower panel of Fig. \ref{fig:erosedcomp} as a dotted line.
It is compatible with the upper limit for optical and near-IR magnitudes corresponding
to the MCPS and VMC detection limit and it shows a much better agreement with 
the observed SED than the standard AGB model. 
In the best-fitting model the condensation temperature  is $T_c=330$ K,
corresponding to an inner radius of the dusty shell of $\sim 5 \times
10^4 \, R_\odot $ or $\sim 3.5 \times 10^{10}$ km. This value is extremely small,
just 10 times bigger than the inner radius of the dusty shell obtained
for the standard model.
This implies that the mass-loss production in J050343.02-664456.7
has dropped to zero extremely recently --of the order of $\sim 100$ years ago
assuming a shell expansion velocity of 10
km s$^{-1}$. 

To conclude, we showed that the observed SED of J050343.02-664456.7 is
not fully compatible with a standard model for AGB stars, showing a
flux excess at wavelengths shorter than $\sim 4$ \micron.  We proposed
that this source could belong to the --rare-- class of objects in
transition from the AGB to the planetary nebula stage.  Further
evidence supporting this idea could be found in the fact that some of
the SEDs of the 13 EROs identified by \cite{grue+2009} show hints of a
secondary peak at near-IR wavelengths.  We hence point out that it
may be not so straightforward to classify all EROs as AGB carbon
stars.

\end{appendix}
%
%
%

\end{document}